# Probability Bracket Notation, Markov Chains, Stochastic Processes, and Microscopic Probabilistic Processes

Xing M. Wang
Sherman Visual Lab, Sunnyvale, CA, USA

## Abstract

Inspired by the Dirac vector probability notation (VPN), we propose the Probability Bracket Notation (PBN), a new set of symbols defined similarly (but not identically) as in the VPN. Applying the PBN to fundamental definitions and theorems for discrete and continuous random variables, we show that the PBN could play a similar role in the probability space as the VBN in the Hilbert vector. Our system P-kets are identified with the probability vectors in Markov chains (MC). The master equation of homogeneous MC in the Schrodinger pictures can be basis-independent. Our system P-bra is linked to the Doi state function and the Peliti standard bra. Transformed from the Schrodinger picture to the Heisenberg picture, the time dependence of the system P-ket of a homogeneous MC (HMC) is shifted to the observable as a stochastic process. Using the correlations established by the special Wick rotation (SWR), the microscopic probabilistic processes (MPPs) are investigated for single and many-particle systems. The expected occupation number of particles in quantum statistics is reproduced by associating time with temperature (the Wick-Matsubara relation).

## 1. Introduction

Dirac notation ([1], §7.2; [2], Appendix I) is a powerful tool to manipulate vectors in the Hilbert space. It has been widely used in Quantum Mechanics (QM) and Quantum Field Theories (QFT). It has also been introduced to Information Retrieval (IR) [2-3]. We call this set of notations the Vector Bracket Notation (VBN)**.** The main beauty of the VBN is that many formulas in QM can be presented in a symbolic abstract way, independent of the state expansion or the basis selection, which, when needed, is easily done by inserting an identity operator.

Inspired by the great success of VBN for vectors in the Hilbert space, we propose the Probability Bracket Notation (PBN), a new set of symbols for probability modeling in the probability sample space. In the time-independent probability sample space, we define symbols like the probability bra (P-bra), P-ket, P-bracket, P-basis, system state P-ket/bra, identity operator, and more in the PBN as their counterparts of the VBN. We demonstrate that the PBN has a similar power as the VBN: various probability formulas can also be presented abstractly, independent of the P-basis.

In terms of the PBN, we identify time-dependent system P-kets with so-called probability vectors ([4], §11.1), which play important roles in Markov chains ([4] §11 and [5]). We show how to construct the left-acting or the right-acting transition of discrete-time Markov chains (MC). This will help us to understand related topics like diffusion maps [8-9] in data clustering [15, 18]. We apply PBN to some important stochastic processes and present the time evolution equation (or the master equation) of the time-continuous homogeneous MC (TC-HMC) in the Schrodinger picture. The system P-bra can be linked to the state function or standard bra introduced in Doi-Peliti Techniques [16-18]. We introduce the Heisenberg picture of stochastic processes and explain the implication of the change from the Schrodinger to the Heisenberg picture. By using the correlations from the special Wick rotation (SWR) [19], in addition to the microscopic probabilistic processes (MPP) of one single particle discussed in [20], the MPP of many-particle systems is investigated, and the expected occupation number of particles is reproduced by associating time with temperature through the Wick-Matsubara relation $it \to t \to \hbar / kT$ .

In Appendix A, we give an item-to-item comparison of PBN and VBN in two short tables, summarizing the similarities and differences of the two notations. In Appendix B, the master equation for the TC-HMC with continuous states is derived.

## 2. PBN and Time-independent Discrete Random Variables

In this section, we introduce the basic symbols of the PBN for a time-independent discrete sample space. We define the probability event-bra (P-bra), probability evidence-ket (P-ket), and their bracket (P-bracket) to represent the conditional probability. In the process, we often use the definitions, theorems, and samples in the book by Grinstead and Snell [4], denoted as "Ref [4]-" or "Based on Ref. [4]". Our definitions, suggestions, and theorems will be stated as propositions. In this article, the bra, ket, and bracket defined in the VBN are named V-bra, V-ket, and V-bracket.

### 2.1. The Basic Symbols of the Probability Bracket Notation

**Definition 2.1.1** (Distribution Function**,** Based on Ref [4]-Definition 2.2): Let $X$ be a random variable that denotes the value of the outcome of a certain experiment, and assume that this experiment has only finitely many possible outcomes. Let Ω be the sample space of the experiment (i.e., the set of all possible values of $X$, or equivalently, the set of all possible outcomes of the experiment). The distribution function for $X$ is a real-valued function $m$ whose domain is Ω and which satisfies:

$$\begin{aligned} &1.\ m(\omega_i) \geq 0, \quad \text{for all } \omega_i \in \Omega, \quad \text{and} \\ &2.\ \sum_{\omega_i \in \Omega} m(\omega_i) = 1. \end{aligned} \qquad (2.1.1)$$

For any subset $E$ of $\Omega$, the probability of E is given by:

$$P(E) = \frac{|E|}{|\Omega|} = \sum_{\omega_i \in E} m(\omega_i) = \sum_{\omega_i \in E} \frac{|\omega_i|}{|\Omega|} \qquad (2.1.2a)$$

**Definition 2.1.2** (Conditional Probability, [4], page 134): For any subset A and B of Ω, the conditional probability of event A under evidence B is defined by:

$$P(A \mid B) = \frac{P(A \cap B)}{P(B)} = \frac{|A \cap B|}{|B|} \qquad (2.1.2b)$$

**Proposition 2.1.1** (Probability event-bra and evidence-ket)**:** Suppose that A and B are subsets of a sample space Ω, we denote:

1. The symbol $P(A| \equiv (A|$ represents a probability event bra, or $P$-bra;
2. The symbol $|B)$ represents a probability evidence ket or $P$-ket.

**Proposition 2.1.2** (Probability Event-Evidence Bracket): The conditional probability of event $A$ given evidence $B$ in the sample space Ω is denoted by the probability event-evidence bracket or P-bracket:

$$P(A \mid B) \equiv \frac{P(A \cap B)}{P(B)} = \frac{|A \cap B|}{|B|}, \text{ if } 0 < \frac{|B|}{|\Omega|} \leq 1 \quad (2.1.3a)$$

By definition, the *P*-bracket has the following properties (see [4], §1.2):

Discrete *RV*: $P(A \mid B) = 1 \;\; if \;\; A \supseteq B \neq \varnothing$ (2.1.4a)

Continuous *RV*: $P(A \mid B) = 1 \;\text{ if }\; A \supset \bar{a} \supset B \neq \varnothing \;\&\; \int_{\bar{a}} dx > 0$ (2.1.4b)

$P(A \mid B) = 0 \quad if \;\; A \cap B = \varnothing$ (2.1.4c)

$P(A \mid B) = P(A \mid \Omega)$, if *A* and *B* are mutually indepedent, see Eq. (2.2.12) (2.1.4d)

These are the most important properties of the P-bracket. We see that the P-bracket is not the inner product of two vectors, while in VBN the V-bracket is.

**Proposition 2.1.3:** For any event *E* in the sample space **Ω**, the probability *P*(*E*) can be written as:

$$P(E) = P(E \mid \Omega) \quad (2.1.5a)$$

Proof: By definition and with Eq. (2.1.2-3), *P*(*E*/**Ω**) can be validated as follows:

$$P(E \mid \Omega) = \frac{|E \cap \Omega|}{|\Omega|} = \frac{|E|}{|\Omega|} \underset{(2.1.2a)}{=} P(E) = \sum_{\omega_i \in E} m(\omega_i) \equiv \sum_{\omega_i \in E} P(\omega_i \mid \Omega) \quad (2.1.5b)$$

The *P*-bracket defined in (2.1.3a) now can also be written as:

$$P(A \mid B) = \frac{P(A \cap B)}{P(B)} = \frac{P(A \cap B \mid \Omega)}{P(B \mid \Omega)} \quad (2.1.3b)$$

Let us use our PBN to rewrite the proof of some theorems in Ref. [4].

**Theorem 1.1.1** (Ref. [4]-Theorem 1.1.4): If *A* and *B* are two disjoint subsets of **Ω**, then:

$$P(A \cup B) = P(A) + P(B) \text{ or } P(A \cup B \mid \Omega) = P(A \mid \Omega) + P(B \mid \Omega) \quad (2.1.6)$$

*Proof in PBN:*

$$P(A \cup B \mid \Omega) = \sum_{\omega \in (A \cup B)} P(\omega \mid \Omega) = \sum_{\omega \in A} P(\omega \mid \Omega) + \sum_{\omega \in B} P(\omega \mid \Omega)$$

$$= P(A \mid \Omega) + P(B \mid \Omega) \quad (2.1.8)$$

**Theorem 2.1.2** (Ref. [4]-Theorem 1.1.2): If $A_1, \ldots, A_n$ are pairwise disjoint subsets of **Ω** (i.e., no two of the $A_i$'s have an element in common), then

$$P(A_1 \cup A_2 \ldots \cup A_n) = \sum_{\mu=1}^{n} P(A_\mu)$$

Proof in PBN**:** By definition (2.1.5b):

$$P(A_1 \cup A_2 \ldots \cup A_n \mid \Omega) = \sum_{\mu=1}^{n} \sum_{\omega \in A_\mu} P(\omega \mid \Omega) = \sum_{\mu=1}^{n} P(A_\mu \mid \Omega) \tag{2.1.9a}$$

**Theorem 2.1.3** (Ref. [4]-Theorem 1.1.5): $P(\tilde{A}) = 1 - P(A)$

Proof in PBN**:** Using definition (2.1.5b), we have:

$$P(\tilde{A} \mid \Omega) = \sum_{\omega \in \tilde{A}} P(\omega \mid \Omega) = \sum_{\omega \in \Omega} P(\omega \mid \Omega) - \sum_{\omega \in A} P(\omega \mid \Omega) \tag{2.1.9b}$$

$$\equiv P(\Omega \mid \Omega) - P(A \mid \Omega) = 1 - P(A \mid \Omega) \tag{2.1.9b}$$

**Theorem 2.1.4:** Bayes formula (see [4], §2.1)

Proof in PBN**:**

$$P(A \mid B) \underset{(2.1.3a)}{=} \frac{|A \cap B|}{|B|} = \frac{|A \cap B|}{|A|} \frac{|\Omega|}{|B|} \frac{|A|}{|\Omega|}$$

$$= P(B \mid A) \frac{1}{P(B \mid \Omega)} P(A \mid \Omega) = \frac{P(B \mid A) P(A)}{P(B)} \tag{2.1.10}$$

By definition 2.1.1, the set of all elementary events ω, associated with the random variable $X$, is the sample space Ω, and they are mutually disjoint:

$$\bigcup_{\omega_i \in \Omega} \omega_i \in \Omega, \quad \omega_i \cap \omega_j = \delta_{ij}\, \omega_i \tag{2.1.11}$$

Using Eq. (2.1.4), we have the following $P$-brackets (orthonormality) for base elements:

$$P(\omega_i \mid \omega_j) = \delta_{ij} \tag{2.1.12}$$

**Proposition 2.1.4** (*Probability Sample Basis and Identity operator*): Like in the Hilbert space, the complete mutually disjoint (CMD) events in (2.1.12) form a probability sample basis (or *P-basis*) of the sample space, associated with the random variable $X$. We can use them to define the following identity operator, which can be inserted into a $P$-bracket:

$$\sum_{\omega \in \Omega} |\, \omega) P(\omega \,| = \sum_{i=1}^{N} |\, \omega_i) P(\omega_i \,| = \hat{I}. \tag{2.1.13}$$

*Proof***:** Suppose that $A$ and $B$ are subsets of a sample space Ω, we have:

$$P(A \mid \hat{I} \mid B) = \sum_{\omega \in \Omega} (A \mid \omega) P(\omega \mid B) \underset{(2.1.4)}{=} \sum_{\omega \in A} P(\omega \mid B)$$

$$\underset{(2.1.5b)}{=} \sum_{\omega\in A} \frac{|\omega\cap B|}{|B|} = \frac{\sum_{\omega\in A\cap B} |\omega\cap B|}{|B|} = \frac{|A\cap B|}{|B|} \equiv P(A\,|\,B) \tag{2.1.14}$$

Now we can expand the system $P$-ket. Its right expansion is:

$$|\,\Omega) = \hat{I}\,|\,\Omega) = \sum_i |\,\omega_i)P(\omega_i\,|\,\Omega) = \sum_i m(\omega_i)\,|\,\omega_i) \tag{2.1.15a}$$

For the system $P$-bra, its left expansion is:

$$P(\Omega\,| = P(\Omega\,|\,\hat{I} = \sum_i^N P(\Omega\,|\,\omega_i)(\omega_i\,| \underset{(2.1.4)}{=} \sum_i^N P(\omega_i\,| \tag{2.1.15b}$$

The two expansions are quite different, although their $P$-bracket is consistent with the requirement of normalization:

$$1 = P(\Omega) \equiv P(\Omega\,|\,\Omega) = \sum_{i,j=1}^N P(\omega_i\,|\,m(\omega_i)\,|\,\omega_j) = \sum_{i,j=1}^N m(\omega_i)\,\delta_{ij} = \sum_{i=1}^N m(\omega_i) \tag{2.1.15c}$$

Here we see the essential difference between a $P$-bra and a $P$-ket. The system $P$-ket, when expended, has the probability for each of its members according to the distribution function. On the other hand, the expansion of a system $P$-bra does not contain any information about the probability distribution. This is a quite different behavior, compared to the V-bra and V-ket in Hilbert space, where one is the Hermitian conjugation of the other. This asymmetry is not only true for the sample space Ω but also true for any subset $E$ in Ω. As a $P$-ket, its right expansion is:

$$|\,E) = \hat{I}\,|\,E) = \sum_i |\,\omega_i)P(\omega_i\,|\,E) \underset{(2.1.10)}{=} \sum_i |\,\omega_i)\frac{P(E\,|\,\omega_i)P(\omega_i\,|\,\Omega)}{P(E\,|\,\Omega)} = \frac{\sum_{\omega\in E}|\,\omega)P(\omega\,|\,\Omega)}{P(E\,|\,\Omega)} \tag{2.1.16a}$$

The $P$-ket $|E)$ gives the base events it contains and the conditional probability of each event under evidence $E$. Note that $P(\Omega\,|\,E) = 1$, as Eq. (2.1.4a) predicted:

$$P(\Omega\,|\,E) = \frac{\sum_{\omega\in E} P(\Omega\,|\,\omega)P(\omega\,|\,\Omega)}{P(E\,|\,\Omega)} = \frac{\sum_{\omega\in E} P(\omega\,|\,\Omega)}{P(E\,|\,\Omega)} = \frac{P(E\,|\,\Omega)}{P(E\,|\,\Omega)} = 1$$

As for a $P$-bra, its left expansion is:

$$P(E\,| = P(E\,|\,\hat{I} = \sum_i P(E\,|\,\omega_i)P(\omega_i\,| = \sum_{\omega\in E} P(\omega_i\,| \tag{2.1.16b}$$

We see that the $P$-bra $P(E|$ only has the basis events it contains. Again, the bracket $P(E|\Omega)$ gives us the right probability value by using Eq. (2.1.5b).

If the distribution function is time-dependent, as in Markov chains (see §4), we will use the system $P$-ket $|\Omega_t) \equiv |\Omega^{(t)}) \equiv |\Omega(t))$ as a probability vector, representing the outcomes and their probabilities at time $t$, while the system $P$-bra $P(\Omega|$ represents the set of all possible outcomes (basis P-bras). Their expansion and normalization are similar to time-independent sample space:

$$|\Omega_t) \equiv |\Omega^{(t)}) = \hat{I}\,|\Omega_t) = \sum_i |\,\omega_i\,)(\omega_i\,|\,\Omega_t) = \sum_{\omega\in\Omega} m(\omega_i, t)\,|\,\omega_i) \tag{2.1.17a}$$

$$P(\Omega| = P(\Omega\,|\,I = \sum_i P(\Omega\,|\,\omega_i)P(\omega_i\,| = \sum_{\omega\in\Omega} P(\omega_i\,| \tag{2.1.17b}$$

$$P(\Omega\,|\,\Omega_t) = \sum_{\omega\in\Omega} P(\Omega\,|\,\omega_i)m(\omega_i, t) = \sum_{\omega\in\Omega} m(\omega_i, t) = 1 \tag{2.1.17c}$$

Note: Eq. (2.1.17c) implies that the system P-ket $|\Omega_t)$ is normalized.

**Proposition 2.1.5** (System P-ket and P-bra): A distribution function represents the state of a system of the sample space. Because the P-ket $|\Omega)$ describes such a system state, we call $|\Omega)$ the system state P-ket (or system P-ket) of the sample space; we call $P(\Omega|$ the system state P-bra (or system P-bra). If an operation involves the knowledge of the distribution function, we should always start with the system P-ket and its expansion.

Using the identity operator of Eq. (2.1.14), the *Bayes formula* (2.1.10) now can be written:

$$P(A\,|\,B) \; = \frac{P(B\,|\,A)P(A\,|\,\Omega)}{P(B\,|\,\Omega)} = \frac{P(B\,|\,A)P(A\,|\,\Omega)}{\sum_i (B\,|\,\omega_i)P(\omega_i\,|\Omega)} \tag{2.1.18}$$

To help readers get familiar with the PBN, here are some simple examples (see [4], §1.2).

**Example 2.1.1** (*Rolling a Die*, Ref. [4]-Example 2.6-2.8): A die is rolled once. We let $X$ denote the outcome of this experiment. Then the sample space for this experiment is the 6-element set $\Omega = \{1, 2, 3, 4, 5, 6\}$. We assumed that the die was fair, so the distribution function is defined by $m(i) = 1/6$, for $i = 1, \ldots, 6$.

Using PBN, we have the identity operator for this sample space:

$$\sum_{i=1}^{6} |\,i)P(i\,| = 1 \tag{2.1.19a}$$

And, because the 6 outcomes have the same probability *p*, we can calculate the probability for each outcome:

$$1 = P(\Omega \mid \Omega) = \sum_{i=1}^{6} P(\Omega \mid i) P(i \mid \Omega) = \sum_{i=1}^{6} P(\, i \mid \Omega) = 6p \qquad (2.1.19b)$$

Hence the probability for each outcome has the same value:

$$P(i) \equiv P(i \mid \Omega) = p = \frac{1}{6} \qquad (2.1.19c)$$

**Example 2.1.2:** (Rolling a Die**,** Ref. [4]-Example 2.8, Example 2.1 continued): If *E* is the event that the result of the roll is an even number, then *E* = {2, 4, 6} and *P*(*E*) = *m*(2) + *m*(4) + *m*(6) =1/6 + 1/6 + 1/6 = ½

Using PBN, the probability of event *E* can be easily calculated using Eqs. (2.1.4-9) as:

$$\begin{aligned} P(E) &\equiv P(E \mid \Omega) = P(E \mid \hat{I} \mid \Omega) = \sum_{i=1}^{6} P(E \mid i) P(i \mid \Omega) \\ &= \sum_{i \in E} P(i \mid \Omega) = \sum_{i=2,4,6} p = 3 \cdot \frac{1}{6} = \frac{1}{2} \end{aligned} \qquad (2.1.19d)$$

Applying the Bayes formula (2.1.10a) to the event *i* and *E* in our die sample space and using Eqs. (2.1.3a), (2.1.7) and (2.1.19), we can easily calculate the conditional probability *P*(*i*|*E*) as follows:

$$P(i \mid E) = \frac{P(E \mid i) P(i \mid \Omega)}{P(E \mid \Omega)} = \begin{cases} \dfrac{P(i \mid \Omega)}{1/2} = \dfrac{1/6}{1/2} = \dfrac{1}{3}, \ (i \ even) \\ 0, \quad (i \ odd) \end{cases} \qquad (2.1.20)$$

The expansions of the system *P*-ket and system *P*-bra of this sample space are quite different:

$$|\, \Omega) = \hat{I} \mid \Omega) = \sum_{i=1}^{6} |\, i) P(i \mid \Omega) = \sum_{i=1}^{6} \frac{1}{6} |\, i) \qquad (2.1.21a)$$

$$P(\Omega \,| = P(\Omega \mid \hat{I} = \sum_{i=1}^{6} P(\Omega \mid i) P(i \,| = \sum_{i=1}^{6} P(i \,| \qquad (2.1.21b)$$

But their *P*-bracket is consistent with the fundamental property:

$$P(\Omega \mid \Omega) = \sum_{i=1}^{6} P(i \,| \sum_{j=1}^{6} \frac{1}{6} |\, j) = \sum_{i,j=1}^{6} \frac{1}{6} \delta_{ij} = 1 \qquad (2.1.21c)$$

Next, let us have a brief discussion on independent events.

**Definition 2.1.2** (Independent events**,** Ref. [4]-Definition 4.1): Let E and F be two events. We say that they are independent if either

1) Both events have positive probability and $P(E|F) = P(E)$ and $P(F|E) = P(F)$, or
2) At least one of the events has zero probability.

**Theorem 2.1.5** (Ref. [4]-Theorem 4.1): Two events $E$ and $F$ are independent if and only if $P(E \cap F) = P(E)\,P(F)$.

Proof in PBN**:** From definition (2.1.3), and Definition 2.1.2, the proof is quite simple:

$$P(E \cap F) \underset{(2.1.3a)}{=} P(E\,|\,F)\cdot P(F) \underset{\text{Def. 2.1.2}}{=} P(E\,|\,\Omega)\cdot P(F\,|\,\Omega) \tag{2.1.22a}$$

$$P(E\,|\,F) \underset{(2.1.3b)}{=} \frac{P(E \cap F\,|\,\Omega)}{P(F\,|\,\Omega)} \underset{(2.1.22a)}{=} \frac{P(E\,|\,\Omega)P(F\,|\,\Omega)}{P(F\,|\,\Omega)} = P(E\,|\,\Omega) \tag{2.1.22b}$$

**Theorem 2.1.6** (Ref. [4], Theorem 1.3)**:** For any complete mutually disjoint (CMD) set $\{H_i\}$ in $\Omega$ and any event $E$ in $\Omega$, we have:

$$P(E\,|\,\Omega) = \sum_i P(E\,|\,H_i)P(H_i\,|\,\Omega) \tag{2.1.23}$$

*Proof in PBN***:** A CMD *set* has the following properties:

$$P(H_i\,|\,H_i) = \delta_{i\,j}, \quad \Omega = \bigcup_j H_j \tag{2.1.24}$$

Using the above properties we have: $E = \sum_i (E \cap H_i) \equiv \sum_i A_i, \quad A_i \cap A_j = \delta_{i\,j}\,A_i$; hence:

$$P(E\,|\,\Omega) \underset{(2.1.9a)}{=} \sum_i P(E \cap H_i\,|\,\Omega) \underset{(2.1.3b)}{=} \sum_i P(E\,|\,H_i)P(H_i\,|\,\Omega) \tag{2.1.25}$$

This implies that the CMD sets $\{H_i\}$ defined in Eq. (2.1.24) construct another identity operator of the sample space, which can be inserted into a $P$-bracket:

$$\hat{I} = \sum_i |\,H_i)P(H_i\,| \tag{2.1.26}$$

**Proposition 2.1.6** (Identity operator from any CMD sets): If sets $\{H_i\}$ are CMD, as in Eq. (2.14), then we can build an identity operator as in Eq. (2.1.26):

$$\sum_i P(A\,|\,H_i)P(H_i\,|\,B) \underset{\substack{Bayes\\formula}}{=} \sum_i \frac{P(H_i\,|\,A)P(A\,|\,\Omega)}{P(H_i\,|\,\Omega)}\,\frac{P(B\,|\,H_i)P(H_i\,|\,\Omega)}{P(B\,|\,\Omega)}$$

$$=\sum_i \frac{P(B|H_i)P(H_i|A)P(A|\Omega)}{P(B|\Omega)} \underset{(2.1.25)}{=} \frac{P(B|A)P(A|\Omega)}{P(B|\Omega)} \underset{\substack{Bayes \\ formula}}{=} P(A|B) \qquad (2.1.27)$$

Using Eq. (2.1.26), the *Bayes formula* in Eq. (2.1.10, 2.1.18) now can be written as:

$$P(A|B) = \frac{P(B|A)P(A|\Omega)}{(B|\Omega)} = \frac{P(B|A)P(A|\Omega)}{\sum_i (B|H_i)P(H_i|\Omega)} \qquad (2.1.28)$$

This is identical to the version in Ref. [4], §2.1.

## 2.2. Observables and their Expectation Values

**Definition 2.2.1** (*Expectation Value*, Ref. [4]-Definition 6.1): Let $X$ be a numerically-valued discrete random variable with a sample space $\Omega$ and a distribution function $m(x)$. The expected value $E(X)$ is defined as

$$E(X) \equiv \sum_{x\in\Omega} x\, m(x) \qquad (2.2.1)$$

provided this sum converges absolutely. If the above sum does not converge absolutely, then we say that $X$ does not have an expected value.

**Proposition 2.2.1** (*Observable and its eigen-ket and eigen-bra*): We call the random variable $X$ of a sample space $\Omega$ an *observable*. The fact that $X$ takes value $x$ at $P$-ket $|x)$ can be denoted in PBN as an operator acting on them:

$$X\,|\,x) = x\,|\,x) \qquad (2.2.2a)$$

According to Proposition 2.1.4, they form a base of $\Omega$ associated with $X$:

$$P(x|x') = \delta_{xx'}, \quad \sum_{x\in\Omega} |\,x)P(x| = 1 \qquad (2.2.2b)$$

Now we have the following compact expression of expectation value.

**Proposition 2.2.2** (*Expectation Value*): The expected value of the observable $X$ in the sample space $\Omega$ can be expressed as:

$$\langle X\rangle \equiv \overline{X} \equiv E(X) = P(\Omega|\,X\,|\Omega) \qquad (2.2.3)$$

*Proof*:

$$P(\Omega|\,X\,|\Omega) = \sum_{x\in\Omega} P(\Omega|\,X\,|\,x)P(x|\Omega) = \sum_{x\in\Omega} P(\Omega|\,x\,|\,x)P(x|\Omega)$$

$$= \sum_{x\in\Omega} P(\Omega \mid x) x P(x \mid \Omega) = \sum_{x\in\Omega} x P(x \mid \Omega) \equiv \sum_{x\in\Omega} x m(x) = E(X) \tag{2.2.4}$$

If *F(X)* is a continuous function of the observable *X*, then it is easy to show that:

$$\langle F(X) \rangle \equiv E(F(X)) \equiv P(\Omega \mid F(X) \mid \Omega) == \sum\nolimits_{x\in\Omega} F(x) m(x) \tag{2.2.5}$$

**Definition 2.2.2** (*Variance***,** Based on Ref. [4]-Definition 6.5): Let *X* be a real-valued random variable with density function *f*(*x*). The variance $\sigma^2 = V(X)$ is defined by

$$\sigma^2 \equiv V(X) \equiv P(\Omega \mid (X - \bar{X})^2 \mid \Omega) \tag{2.2.6}$$

It can be easily seen that:

$$\begin{aligned} \sigma^2 &= P(\Omega \mid (X^2 - 2X\bar{X} + \bar{X}^2) \mid \Omega) = \langle X^2 - 2X\bar{X} + \bar{X}^2 \rangle \\ &= \langle X^2 \rangle - 2\bar{X}\langle X \rangle + \bar{X}^2 = \langle X^2 \rangle - \bar{X}^2 \end{aligned} \tag{2.2.7}$$

**Example 2.2.1** (*Rolling a Die***,** Example 2.1.1 continued): We have the following observable in the die sample space, as based on Eq. (2.1.19),

$$X \mid i) = i \mid i), \quad i \in \{1,2,...6\} \tag{2.2.8a}$$

Its expectation value can be readily calculated:

$$\begin{aligned} P(\Omega \mid X \mid \Omega) &= \sum_{i=1}^{6} P(\Omega \mid X \mid i) P(i \mid \Omega) = \sum_{i=1}^{6} P(\Omega \mid i \mid i) P(i \mid \Omega) \\ &= \sum_{i=1}^{6} (\Omega \mid i) \frac{i}{6} = \sum_{i=1}^{6} \frac{i}{6} = \frac{21}{6} = \frac{7}{2} \end{aligned} \tag{2.2.8b}$$

The variance can be calculated as:

$$\sigma^2 = \langle X^2 \rangle - \bar{X}^2 = \sum_{i=1}^{6} \frac{1}{6} i^2 - \frac{49}{4} = \frac{91}{6} - \frac{49}{4} = \frac{45}{4} \tag{2.2.8c}$$

**Definition 2.2.3** (*Conditional Expectation Value***,** based on Ref. [4]-Definition 6.2): If *F* is any event and *X* is a random variable with sample space = $\{x_1, x_2, \ldots\}$, then the conditional expectation given *F* is defined by

$$E[X \mid F] \equiv \sum_j x_j P(X = x_j \mid F) = \sum_j x_j P(x_j \mid F) \tag{2.2.9}$$

**Proposition 2.2.2:** In PBN, we can express conditional variance (2.2.6) as:

$$E[X \mid F] = P(\Omega \mid X \mid F) \tag{2.2.10}$$

*Proof:*

$$P(\Omega \mid X \mid F) = \sum\nolimits_x P(\Omega \mid X \mid x) P(x \mid F) = \sum\nolimits_x P(\Omega \mid x \mid x) P(x \mid F)$$
$$= \sum\nolimits_x P(\Omega \mid x)\, x\, P(x \mid F) = \sum\nolimits_x x\, P(x \mid F) = \sum\nolimits_j x_j\, P(x_j \mid F) \tag{2.2.11}$$

The conditional expectation is used mostly in the form provided by the following theorem.

**Theorem 2.2.3** (Ref. [4]-Theorem 6.5): Let $X$ be a random variable with sample space. If $F_1, F_2, \ldots, F_r$ are events such that $F_i \cap F_j = \emptyset$; for $i \neq j$ and $\Omega = \bigcup_j F_j$, then

$$E[X] = \sum_j E[X \mid F_j] P(F_j) \tag{2.2.12}$$

Proof in PBN: Using Eq. (2.2.11), (2.1.26), and (2.2.4), we have:

$$\sum_j E[X \mid F_j] P(F_j) \underset{(2.2.10)}{=} \sum_j P(\Omega \mid X \mid F_j) P(F_j \mid \Omega)$$
$$\underset{(2.2.11)}{=} \sum_k \sum_j x_k P(x_k \mid F_j) P(F_j \mid \Omega) \underset{(2.1.26)}{=} \sum_k x_k P(x_k \mid \Omega) \underset{(2.2.1)}{=} \langle X \rangle$$

**Example 2.2.4** (Rolling Two Die): We have two observables in the 2-die sample space, as the extension of Eqs. (2.1.19),

$$X \mid i, j) = i \mid i, j), \quad Y \mid i, j) = j \mid i, j), \quad i, j \in \{1,2,...6\} \tag{2.2.8a}$$

Its expectation value of $X{\cdot}Y$ can be readily calculated:

$$P(\Omega \mid X \cdot Y \mid \Omega) = \sum_{i,j=1}^{6} P(\Omega \mid X \cdot Y \mid i, j) P(i, j \mid \Omega) = \sum_{i,j=1}^{6} P(\Omega \mid i \cdot j \mid i, j) P(i, j \mid \Omega)$$
$$= \sum_{i,j=1}^{6} P(\Omega \mid i, j) \frac{i \cdot j}{36} = \left( \sum_{i=1}^{6} \frac{i}{6} \right)^2 = \left( \frac{7}{2} \right)^2 = \frac{49}{4} \tag{2.2.8b}$$

## 2.3. Independent Discrete Random Variables

**Definition 2.3.1** Joint random variable**,** based on Ref. [4]-Definition 4.3): Let $X_1, X_2, \ldots, X_n$ be random variables associated with an experiment. Suppose that the sample space (i.e., the set of possible outcomes) of $X_i$ is the set $\Omega_i$. Then the joint random variable (or random vector) $\vec{X} = (X_1, X_2, \ldots, X_n)$ is defined to be the random variable whose outcomes consist of ordered $n$-tuples of outcomes, with the $i$th coordinate lying in the set $\Omega_i$. The sample space of $\vec{X}$ is the Cartesian product of the $\Omega_i$'s:

$$\Omega = \Omega_1 \otimes \Omega_2 \otimes \cdot \cdot \cdot \otimes \Omega_n. \tag{2.3.1}$$

The joint distribution function of $\vec{X}$ is the function that gives the probability of each of the outcomes of $\vec{X}$ .

**Proposition 2.3.1:** In PBN, the sample space of the joint variable $\vec{X}$ is written as:

$$|\,\Omega) = \prod_{i=1}^{n} |\,\Omega_i) \tag{2.3.2a}$$

The factor sample space $|\Omega_i)$ has the following properties:

$$P(\Omega_i\,|\,\Omega_i) = 1, \quad |\,\Omega_i)\,|\,\Omega_j) = |\,\Omega_j)\,|\,\Omega_i), \quad P(\Omega_i\,|\,P(\Omega_j\,| = P(\Omega_j\,|\,P(\Omega_i\,| \tag{2.3.2b}$$

The base event ket of joint random variables can be written as:

$$|\,r_1, r_1, \ldots, r_n) = \prod_{i=1}^{n} |\,r_i)_i, \quad |\,r_i)_i\,|\,r_j)_j = |\,r_j)_j\,|\,r_i)_i \tag{2.3.2c}$$

This expression is similar to its counterpart in the Fock space ([1], §22.1).

**Definition 2.3.2** (Independent Random Variables, based on Ref. [4]-Definition 4.4): The random variables $X_1, X_2, \ldots, X_n$ are mutually independent if for any choice of $r_1, r_2, \ldots, r_n$:

$$P(X_1 = r_1, X_2 = r_2, \ldots, X_n = r_n) = \prod_{i=1}^{n} P(X_1 = r_1) \tag{2.3.4}$$

Thus, if $X_1, X_2, \ldots, X_n$ are mutually independent, then the joint distribution function of the random vector $\vec{X} = (X_1, X_2, \ldots, X_n)$ is just the product of the individual distribution functions. When two random variables are mutually independent, we shall say more briefly that they are independent.

**Proposition 2.3.3:** In PBN, using proposition 2.3.1, Eq. (2.3.4a) can be written as:

$$P(r_1, r_1, \ldots, r_n\,|\,\Omega) = \prod_{i=1}^{n} {}_i P(r_i\,|\,\Omega_i) = \prod_{i=1}^{n} m_i(r_i) \tag{2.3.5}$$

From Eq. (2.3.5), we can derive the following properties for independent random variables:

$$P(r_1\,|\,\Omega) = \sum_{r_2 \ldots r_n} P(r_1, r_2, \ldots, r_n\,|\,\Omega) = P_1(r_1\,|\,\Omega_1) \prod_{i=2}^{n} \sum_{r_i} P_i(r_i\,|\,\Omega_i) = P_1(r_1\,|\,\Omega_1) \tag{2.3.6}$$

$$P(r_i, r_j \mid \Omega) = P(r_i \mid \Omega) P(r_j \mid \Omega) = P_i(r_i \mid \Omega_i) \cdot P_j(r_j \mid \Omega_j) \quad (2.3.7a)$$

$$P(r_i \mid r_j) = \frac{P(r_i, r_j \mid \Omega)}{P(r_j \mid \Omega)} = \frac{P(r_i \mid \Omega) P(r_j \mid \Omega)}{P(r_j \mid \Omega)} = P(r_i \mid \Omega) = P_i(r_i \mid \Omega_i) \quad (2.3.7b)$$

Eq. (2.3.7) is equivalent to Eq. (2.1.11).

As observables, we have the following eigen-bras and eigen-kets:

$$X_i \mid r_1, \ldots, r_n) = r_i \mid r_1, \ldots, r_n) \quad (2.3.8)$$

Hence, for analytical functions $F_i(x)$ and $k(x)$, we have

$$\langle \sum c_i F_i(X_i) \rangle \equiv P(\Omega \mid \sum c_i F_i(X_i) \mid \Omega) = \sum c_i F_i(\langle X_i \rangle) \quad (2.3.8a)$$

$$\langle \prod_i X_i^{k(i)} \rangle \equiv P(\Omega \mid \prod_i X_i^{k(i)} \mid \Omega) = \prod_i \langle X_i \rangle^{k(i)} \quad (2.3.8b)$$

Here, we have used the following expectation value:

$$\langle X_i \rangle \equiv P(\Omega \mid X_i \mid \Omega) = P(\Omega_i \mid X_i \mid \Omega_i) \quad (2.3.9)$$

Example 2.2.2 (rolling two die) is an example of two independent random variables $X$ and $Y$. Using Eq. (2.3.8b), we can easily recalculate (2.2.8b) as follows:

$$P(\Omega \mid X \cdot Y \mid \Omega) = \langle X \rangle \cdot \langle Y \rangle == \left(\frac{7}{2}\right)^2 = \frac{49}{4} \quad (2.3.10)$$

## 3. PBN and Time-independent Continuous Random Variables

In this section, we define PBN for a *continuous* sample space. Because many formulas read similarly, we only discuss a few selected concepts.

### 3.1. The Sample-basis and the Identity Operator

**Definition 3.1.1** (Continuous Distribution or Density Function)**:** Let $X$ be a random variable that denotes the value of the outcome of a certain experiment, and assume that this experiment has only finitely many possible outcomes. Let Ω be the sample space of the experiment (i.e., the set of all possible values of $X$, or equivalently, the set of all possible outcomes of the experiment.) A distribution function for $X$ is a real-valued function whose domain is Ω and which satisfies:

1. $f(x) \ge 0, \quad for\ all\ x \in \Omega, \quad and$
2. $\int_{x\in\Omega} dx\, f(x) = 1.$ (3.1.1)

For any subset *E* of *Ω*, we define the probability of *E* to be the number *P*(*E*) given by:

$$P(E) = \frac{|E|}{|\Omega|} = \int_{x\in E} dx\, f(x) \qquad (3.1.2)$$

Suppose that *A* and *B* are subsets of a sample space Ω. The conditional probability of event *A* under evidence *B* in sample space Ω, and its properties are the same as in discrete sample space. The theorems 2.1.1-2.1.4 can be proved similarly for continuous sample space. Note the condition in Eq. (2.1.4b):

Continuous RV: $P(A\,|\,B) = 1 \text{ if } A \supset \overline{a} \supset B \neq \varnothing \ \&\ \int_{\overline{a}} dx > 0$ (2.1.4b)

The last condition is needed. Otherwise, we may have the Dirac delta function, as shown in Eq. (3.1.3a) below. For orthonormality, we extend Eq. (2.1.12) to the continuous case:

$$P(x\,|\,x') = \delta(x - x') \qquad (3.1.3a)$$

Then, like in a discrete sample space, these events form *the basis* of the sample space. They can be used to define an identity operator in the sample space:

$$\int_{x\in\Omega} dx\,|\,x)\,P(x\,| = \hat{I} \qquad (3.1.3b)$$

The definition in Eq. (2.1.3a), $P(A\,|\,B) = \dfrac{P(A\cap B)}{P(B)}$ requires *P*(*B*) > 0. But from Eq. (3.1.2), we have *P*(*B*) = 0, if $B = x \in \Omega$. This does not cause problems now, because:

$$P(A\,|\,x) = \frac{P(A\cap x\,|\,\Omega)}{P(x\,|\,\Omega)} = \frac{P(x\,|\,\Omega)}{P(x\,|\,\Omega)} = 1, \text{ if } x \in A \qquad (3.1.3c)$$

**Proposition 3.1.1**: In PBN, the distribution function (or probability density) is denoted by:

$$f(x) \equiv P(x\,|\,\Omega) \qquad (3.1.4a)$$

The proof is similar to that in the discrete sample case. We check here if it is consistent with the normalization requirement:

$$P(\Omega\,|\,\Omega) = \int P(\Omega\,|\,x)\,dx\,P(x\,|\,\Omega) = \int_{x\in\Omega} dx\,P(x\,|\,\Omega) = \int_{x\in\Omega} dx\, f(x) = 1 \qquad (3.1.4b)$$

**Example 3.1.1** (*Darts*, based on Ref. [4], Example 3.8-2.9): A game of darts involves throwing a dart at a circular target of unit radius. Suppose we throw a dart once so that it hits the target, and we observe where it lands. To describe the possible outcomes of this experiment, it is natural to take the set of all the points in the target as our sample space. It is convenient to describe these points by their rectangular coordinates, relative to a coordinate system with the origin at the center of the target so that each pair $(x, y)$ of coordinates with $x^2 + y^2 \leq 1$ describes a possible outcome of the experiment. Then $\Omega = \{(x, y): x^2 + y^2 \leq 1\}$ is a subset of the Euclidean plane, and the event $E = \{(x, y): y > 0\}$, for example, corresponds to the statement that the dart lands in the upper half of the target, and so forth.

Assuming a uniform distribution, the probability of the event that the dart lands in any subset $E$ of the target should be determined by what fraction of the target area lies in $E$. Thus, we can calculate $P(E|\,\Omega)$:

$$P(E\,|\,\Omega) = \frac{|\,Area\ of\ E\,|}{|\,Area\ of\ \Omega\,|} = \frac{|\,Area\ of\ E\,|}{\pi} = \int\limits_{x,y\in E} dx\,dy\,f(x,y)$$
$$= \int\limits_{E} dxdy\,f(x,y) \;=\; f(0,0)\int\limits_{E} dxdy = f(0,0)\cdot|\,Area\ of\ E\,| \qquad (3.1.5a)$$

Hence, we get the density function:

$$P(x, y\,|\,\Omega) \equiv f(x,y) = f(0,0) = \frac{1}{\pi} \qquad (3.1.5b)$$

In the continuous probability, the Bayes formula (2.1.10) now can be written:

$$P(A\,|\,B) = \frac{P(B\,|\,A)P(A\,|\,\Omega)}{P(B\,|\,\Omega)} = \frac{P(B\,|\,A)\int\limits_{x\in A} dx\,P(x\,|\,\Omega)}{\int\limits_{x\in B} dx\,P(x\,|\,\Omega)} \qquad (3.1.6)$$

Note it is valid even for singular cases like:

$$P(x\,|\,x') = \frac{P(x'\,|\,x)P(x\,|\,\Omega)}{P(x'\,|\,\Omega)} = \frac{\delta(x-x')P(x\,|\,\Omega)}{P(x'\,|\,\Omega)} = \delta(x-x')$$

**Definition 3.1.2** (Conditional Continuous Density Function**,** Ref. [4], §4.2)

$$f(x\,|\,E) = \begin{cases} \dfrac{f(x)}{P(E)} & if\ \ x \in E \\ 0 \ \ if\ \ x \notin E \end{cases} \qquad (3.1.7)$$

**Proposition 3.1.2**: Using PBN, we can denote the conditional density function as:

$$f(x \mid E) \equiv P(x \mid E) \tag{3.1.8}$$

*Proof:* Using Bayes formula (2.1.10), we have:

$$P(x \mid E) = \frac{P(E \mid x)P(x \mid \Omega)}{P(E \mid \Omega)} = \begin{cases} \dfrac{P(x \mid \Omega)}{P(E \mid \Omega)} \equiv \dfrac{f(x)}{P(E)}, & if \;\; x \in E \\ 0, & if \;\; x \notin E \end{cases}$$

**Proposition 3.1.3** (Conditional Probability of Event E given F**,** see Ref. [4], §4.2):

$$P(F \mid E) = \int_{\Omega} P(F \mid x) dx\, P(x \mid E) = \int_{x \in F} dx\, P(x \mid E) \tag{3.1.9}$$

Proof: We can check that this is consistent with our definition of conditional probability, Eq. (2.1.3a):

$$P(F \mid E) = \int_{x \in F} dx\, P(x \mid E) = \int_{x \in E \cap F} dx \frac{P(x \mid \Omega)}{P(E \mid \Omega)} = \frac{P(E \cap F \mid \Omega)}{P(E \mid \Omega)} \tag{3.1.10}$$

**Example 3.1.2** (Darts, Based on Ref. [4], Example 5.19): In the dart game (cf. Example 3.8, our Example 3.1.1), suppose we know that the dart lands in the upper half of the target. What is the probability that its distance from the center is less than ½?
Here $E = \{(x, y): y \geq 0\}$, and $F = \{(x, y): x^2 + y^2 < (1/2)^2\}$. Hence,

$$P(E \mid \Omega) \equiv P(E) = \frac{|E|}{|\Omega|} = \frac{\pi / 2}{\pi} = \frac{1}{2} \tag{3.1.11a}$$

$$P(F \mid E) = \frac{P(E \cap F \mid \Omega)}{P(E \mid \Omega)} = \frac{|E \cap F|}{|E|} = \frac{(1/2)(\pi/4)}{\pi/2} = \frac{1}{4} \tag{3.1.11b}$$

Here again, the size of F∩E is 1/4 the size of E. The conditional density function is:

$$f(x, y \mid E) \equiv P(x, y \mid E) = \begin{cases} \dfrac{P(x, y \mid \Omega)}{P(E \mid \Omega)} = \dfrac{1/\pi}{1/2} = \dfrac{2}{\pi}, \text{if } (x, y) \in E \\ 0, \quad \text{if } (x, y) \notin E \end{cases} \tag{3.1.11c}$$

**Example 3.1.3** (Exponential Density**,** Ref. [4]-Example 3.17): There are many occasions where we observe a sequence of occurrences, which occur at "random" times. For example, we might be observing emissions of a radioactive isotope, or cars passing a milepost on a highway, or light bulbs burning out. In such cases, we might define a random variable $X$ to denote the time between successive occurrences. Clearly, $X$ is a continuous random variable whose range consists of non-negative real numbers. It is often the case that we can model $X$ by using the exponential density. This density is given by the formula:

$$f(t)=\begin{cases}\lambda e^{-\lambda t}, & if\ t\geq 0\\ 0, & if\ t<0\end{cases} \tag{3.1.12a}$$

Using PBN, we have the following base:

$$P(t\,|\,t')=\delta(t-t'),\quad \int_0^\infty dt\,|\,t)P(t\,|=1,\quad X\,|\,t)=t\,|\,t),\quad P(t\,|\,\Omega)=f(t) \tag{3.1.12b}$$

We can see that the sample space is normalized:

$$P(\Omega\,|\,\Omega)=\int_0^\infty P(\Omega\,|\,t)dt\,P(t\,|\,\Omega)=\int_0^\infty dt\,P(t\,|\,\Omega)=1 \tag{3.1.13}$$

**Example 3.1.4** (Exponential Density**,** Ref. [4]-Example 5.20): We return to the exponential density (cf. Example 3.17). Suppose we are observing a lump of plutonium-239. Our experiment consists of waiting for an emission, promptly starting a clock, and recording the time interval $X$ that passes until the next emission. Experience has shown that $X$ has an exponential density with some parameter λ, which depends upon the size of the lump. Suppose that when we perform this experiment, we notice that the clock reads $r$ seconds, and is still running. What is the probability that an emission happens in a further $s$ seconds?

Let $G(t)$ be the probability that the next particle is emitted after time t. Then

$$G(t)\equiv P(G\,|\,\Omega)=\int_0^\infty P(G\,|\,t')dt'\,P(t'\,|\,\Omega)=\int_{t'\in G}dt'\,f(t') \tag{3.1.14}$$
$$=\int_t^\infty dt'\,\lambda e^{-\lambda t'}=-e^{-\lambda t'}\Big|_t^{\infty}=e^{-\lambda t}$$

Let $E$ be the event "the next particle is emitted after time $r$" and $F$ be the event "the next particle is emitted after time $r + s$." Then

$$P(F\,|\,E)=\frac{P(E\cap F\,|\,\Omega)}{P(E\,|\,\Omega)}=\frac{G(r+s)}{G(r)}=e^{-\lambda s} \tag{3.1.15}$$

## 3.2. The Expectation Value of Continuous Random Variables

**Definition 3.2.1** (Expectation Value**,** Ref. [4]-Definition 6.4)**:** Let $X$ be a numerically-valued continuous random variable with sample space and distribution function $f(x)$. The expected value $E(X)$ is defined as

$$E(X) \equiv \int\limits_{x\in\Omega} dx\, x\, f(x) \qquad (3.2.1)$$

provided this integral converges absolutely. If the above integral does not converge absolutely, then we say that *X* does not have an expected value.

As in the discrete case, the property that random variable *X* takes value *x* in evidence |*x*) can be written in PBN as follows:

$$X \mid x) = x \mid x) \qquad (3.2.2)$$

As in §2.2, *X* is an observable of the sample space, and |*x*) is one of its eigen-kets.

The expression of the expected value of continuous *X* is identical to the discrete case:

$$\langle X \rangle \equiv \bar{X} \equiv E(X) \equiv P(\Omega \mid X \mid \Omega) \qquad (3.2.3)$$

We can see its consistency with the definition (3.2.1):

$$\begin{aligned} &P(\Omega \mid X \mid \Omega) = \int_{x\in\Omega} P(\Omega \mid X \mid x) dx\, P(x \mid \Omega) \\ &= \int_{x\in\Omega} P(\Omega \mid x)\, x\, dx\, P(x \mid \Omega) = \int_{x\in\Omega} dx\, x\, f(x) \equiv E(X) \end{aligned} \qquad (3.2.4)$$

If *F(X)* is a continuous function of observable *X*, then we also have Eq. (2.2.5). The equations (2.2.6-1.2.7) related to Variance are also valid.

**Example 3.2.1** (Darts, see Example 3.1.1): We have two observables:

$$X \mid x, y) = x \mid x, y), \quad Y \mid x, y) = y \mid x, y) \qquad (3.2.5a)$$

Their expectation values can be easily calculated:

$$\begin{aligned} &P(\Omega \mid X \mid \Omega) = \int_{(x,y)\in\Omega} P(\Omega \mid X \mid x, y) dydy\, P(x, y \mid \Omega) \\ &= \int_{(x,y)\in\Omega} P(\Omega \mid X \mid x, y) dydy\, P(x, y \mid \Omega) = \frac{1}{\pi}\int_{(x,y)\in\Omega} xdxdy = 0 \end{aligned} \qquad (3.2.5b)$$

In the last step, we used the property of the integral of an odd function in the symmetric boundary.

**Example 3.2.2** (Expected Life-Time): Let us consider Example 3.1.1 as the light bulbs burning out problem. Then the expected value of the time will be the average lifetime of a bulb:

$$P(\Omega \mid X \mid \Omega) = \int_{t\in\Omega} P(\Omega \mid X \mid t)dt\, P(t \mid \Omega)$$

$$= \int_{t\in\Omega} t\, dt\, f(t) = \int_{t}^{\infty} tdt\lambda\, e^{-\lambda t} = \frac{1}{\lambda} \tag{3.2.6}$$

### 3.3. The Phase Space and the Partition Function of Ideal Gas

The formulas related to continuous random variables are similar to the discrete random variables discussed in §2.3. Now we use the PBN to discuss a system of $N$ non-interacting indistinguishable molecules (see [6], §10.1). The distribution density of a single particle depends on its energy (as in [6], Eq. (10.1b) and (10.9)):

$$f(\varepsilon) = P(x, y, z, p_x, p_y, p_z \mid \Omega_1) = P(\vec{x}, \vec{p} \mid \Omega_1) = f\left(\frac{\vec{p}^2}{2m}\right) = \frac{e^{-\frac{\varepsilon}{kT}}}{z} \equiv \frac{e^{-\beta\varepsilon}}{z} \tag{3.3.1}$$

$$z \equiv \frac{1}{h^3}\int d^3x\, d^3p\, e^{-\beta\varepsilon} = \frac{V}{h^3}\int d^3p\, e^{-\frac{\beta p^2}{2m}} = V\left(\frac{2\pi m}{h^2\beta}\right)^{3/2} \tag{3.3.2}$$

The expectation value of the energy of a single particle is given by:

$$\begin{aligned}\langle\varepsilon\rangle &= P(\Omega_1 \mid \varepsilon \mid \Omega_1) = \int P(\Omega_1 \mid \varepsilon \mid \vec{x}, \vec{p})d^3x\, d^3p\, P(\vec{x}, \vec{p} \mid \Omega_1) = \int \varepsilon\, d^3x\, d^3p\, f(\varepsilon) \\ &= \int d^3x\, d^3p\, (-\frac{\partial}{\partial\beta} e^{-\beta\varepsilon})\frac{1}{z} = -\frac{\partial}{z\partial\beta} z = -\frac{\partial}{\partial\beta}\ln z = \frac{3}{2\beta} = \frac{3kT}{2}\end{aligned} \tag{3.3.3}$$

Because the $N$ particles are indistinguishable, using Eq. (2.3.8a), we have the expectation value of total energy as:

$$\langle\sum_{i=1}^{N}\varepsilon_i\rangle = N\, P(\Omega_1 \mid \varepsilon_1 \mid \Omega_1) = \frac{3NkT}{2} \tag{3.3.4}$$

## 4. Probability Vectors and Markov Chains

### 4.1. Time-Dependent Probability Vectors in a Sample Space

In Ref. [3], we have discussed the normalized Weighted Term Space, which is a bounded $N$-dimensional continuous space over the field of [0, 1], restricted in the unit cube:

$$\mid w\rangle = \sum_{i=1}^{N} w_i \mid k_i\rangle\,, \quad w_i \in [0,1] \subset \Re \tag{4.1.1}$$

These vectors are normalized as:

$$\langle w \mid w \rangle = \sum_{i=1}^{N} w_i^{\ 2} = 1 \tag{4.1.2}$$

In probability theory, we will face another kind of vectors – Probability Vectors (see [4], §11.1). There are probability row vectors (PRV) and probability column vectors (PCV), defined in sample space, restricted in the same unit cube as in Eq. (4.1.1), but is normalized as follows:

$$w_i = P(\omega_i \mid \Omega), \quad \sum_{i=1}^{N} P(\omega_i \mid \Omega) = 1, \quad P(\omega_i \mid \Omega) \geq 0 \tag{4.1.3}$$

Because of this normalization requirement, probability vectors do not form a closed vector space, nor do they form vectors in Hilbert space. But time-dependent probability vectors have important applications in probability theories like the Markov chain ([4], chapter 11; [5]) and the IR models like diffusion maps [8-9], we want to show how to build time-dependent probability vectors from a sample space with a time-dependent distribution function, as described by Eq. (2.1.17). From the system *P*-ket of Eq. (2.1.17a), we can easily form a PCV by mapping ket-to-ket and bra-to-bra as follows:

$$| \Omega_t \rangle = I \, | \Omega_t \rangle = \sum_{i}^{N} | \omega_i \rangle \langle \omega_i \mid \Omega_t \rangle = \sum_{i}^{N} m(\omega_i, t) \, | \omega_i \rangle = \begin{pmatrix} m(\omega_1, t) \\ m(\omega_2, t) \\ \vdots \\ m(\omega_N, t) \end{pmatrix} \tag{4.1.4a}$$

Its counterpart, a row vector, can also be mapped from the system *P*-bra $(\Omega|$ in Eq. (2.1.17b) as:

$$\langle \Omega \,| = \sum_{i}^{N} \langle \omega_i \,| = \begin{bmatrix} 1, & 1, & \cdots, & 1 \end{bmatrix} \tag{4.1.4b}$$

It is time-independent and is not a PRV, because it does not satisfy Eq. (4.1.3). But, from Eq. (4.1.4a-b) and (2.1.17c), we see the normalization is correct:

$$\langle \Omega | \Omega_t \rangle = \sum_{i,j}^{N} \langle \omega_i \,| m(\omega_i, t) \,| \, \omega_j \rangle = \sum_{i}^{N} m(\omega_i, t) = 1 \tag{4.1.5}$$

As we will see in the next section, the transition matrix of a Markov chain is acting on a row vector from the right. We need a PRV with a time-dependent distribution function, which can be obtained as the transpose of PCV in (4.1.4a):

$$\langle \Omega_t \,| = \sum_{i}^{N} m(\omega_i, t) \langle \omega_i \,| = [m(\omega_1, t), m(\omega_2, t), \ldots, m(\omega_N, t)] \tag{4.1.7}$$

It is correctly normalized because of Eq. (2.1.17). We want to point out, that the inner product of (Eq. 4.1.4a) and (4.1.7) usually is not equal to 1, and it does not have any meaning in terms of probability, since:

$$\langle \Omega_t | \Omega_t \rangle = \sum_i^N (m(\omega_i, t))^2 \le 1 \qquad (4.1.8)$$

In summary, to build PCV and PRV in a sample space, we first expand the system P-ket $|\Omega)$ to get the PCV, and then we make a transpose to get the PRV. We will provide more detail on applying probability vectors in the next section.

### 4.2. The Left-acting Transition Operator and Markov Chains

In this section, we briefly discuss the time-discrete and state-discrete Markov chains of discrete state spaces (see [4] Chap.11 or [5]). Our goal is to demonstrate how to use PBN to describe the sample space of a Markov chain. In our discussion, we assume our Markov chain is time-homogeneous.

We also assume our sample space has the following discrete *P*-basis:

$$P(i \mid j) = \delta_{ij}, \quad \sum_{i=1}^{r} |\, i) P(i \,| = I \qquad (4.2.1)$$

The transition matrix element $P_{ij}$ is defined by the transition probability from state i to state j at time t (an integer, measuring steps). Its elements are non-negative real numbers and are time-independent for a time-homogeneous Markov chain [4, 8]:

$$P_{ij} \equiv P(X_{t+1} = j \mid X_t = i) \equiv P(X_{t+1} = j \mid X_t = i) \equiv P(j, t+1 \mid i, t) \qquad (4.2.2a)$$

$$\sum_{j=1}^{r} P_{ij} = 1 \qquad (4.2.2b)$$

In matrix form, if we define a probability row vector (PRV) at t = 0 as $u^{(0)}$, then $P$ acting on the PRV from right $n$ times gives the PRV at time = $n$ ([4], theorem 11.2):

$$u^{(n)} = u^{(0)} P^n, \text{ or: } \quad u^{(n)}{}_i = \sum_j u^{(n)}{}_j \, P^n{}_{ji} \qquad (4.2.2c)$$

**Example 4.2.1** (The Land of Oz*,* Based on Ref. [4], example 11.1-11.3): According to Kemeny, Snell, and Thompson [7], the Land of Oz is blessed by many things, but not by good weather. They never have two nice days in a row. If they have a nice day, they are just as likely to have snow as rain the next day. If the day has snow or rain, they have an even chance of having the same the next day. If there is a change from snow or rain, only half the time is this a change to a nice day. With this information, we form a Markov chain as follows. We take as states the kinds of weather R, N, and S. From the above information we determine the transition probabilities. These are most conveniently represented in a square array as

$$\mathbf{P} = \begin{array}{c} \\ \mathrm{R} \\ \mathrm{N} \\ \mathrm{S} \end{array} \begin{array}{c} \begin{array}{ccc} \mathrm{R} & \mathrm{N} & \mathrm{S} \end{array} \\ \begin{pmatrix} 1/2 & 1/4 & 1/4 \\ 1/2 & 0 & 1/2 \\ 1/4 & 1/4 & 1/2 \end{pmatrix} \end{array} .$$

Let the initial probability vector u equal (1/3, 1/3, 1/3). Then we can calculate the distribution of the states after three days using Theorem 11.2 and our previous calculation of P3. We obtain

$$\begin{aligned} \mathbf{u}^{(3)} = \mathbf{u}\mathbf{P}^3 &= \begin{pmatrix} 1/3, & 1/3, & 1/3 \end{pmatrix} \begin{pmatrix} .406 & .203 & .391 \\ .406 & .188 & .406 \\ .391 & .203 & .406 \end{pmatrix} \\ &= \begin{pmatrix} .401, & .198, & .401 \end{pmatrix} . \end{aligned}$$

In Eq. (4.2.2c) and the above example, the transition matrix acts on the row vector to its left side. This implies that we need a left-acting operator and a PRV.

To use our P-basis, we define the following transition operator in sample space based on the elements $\{p_{ij}\}$:

$$\hat{P} = \sum_{i',j'=1}^{r} | i')P_{i'j'} P(j' | \tag{4.2.2d}$$

The matrix element of the operator in the *P*-basis is the element of the transition matrix:

$$P(i | \hat{P} | j) = \sum_{i',j'=1}^{r} P(i | i')p_{i'j'}P(j' | j) = \sum_{i',j'=1}^{r} \delta_{i'i'} p_{i'j'} \delta_{j'j'} = p_{ij} \tag{4.2.2e}$$

As discussed in §4.3, to build a PRV, we start with a PCV. Because we are talking about the time-evolution of states, the distribution function now is time-dependent, so we use the following system *P*-ket as in Eq. (2.1.17):

$$| \Omega_t ) = | \Omega^{(t)} ) = \sum_{i}^{r} | i)P(i | \Omega^{(t)} ) = \sum_{i}^{r} \omega^{(t)}{}_i \, | i) \qquad \sum_{i}^{r} \omega^{(t)}{}_i = 1 \tag{4.2.3a}$$

Now we build the V-base from the *P*-basis by one-to-one-map:

$$\langle i | j \rangle = \delta_{ij}, \quad \sum_{i=1}^{r} | i \rangle\langle i | = I \tag{4.2.3b}$$

From Eq. (4.2.3) and Eq. (4.1.7), we obtain the PRV as:

$$\langle \Omega^{(t)} | = \sum_{i}^{r} \omega^{(t)}{}_{i} \langle i | = [\omega^{(t)}{}_{1}, \omega^{(t)}{}_{2}, \cdots, \omega^{(t)}{}_{r}], \qquad \sum_{i}^{r} \omega^{(t)}{}_{i} = 1 \tag{4.2.4}$$

The *left-acting* operator, to act on a PRV from right, is defined as

$$\overleftarrow{P} = \sum_{i',j'=1}^{r} | i' \rangle p_{i'j'} \langle j' | \tag{4.2.5a}$$

**Proposition 4.2.1** (*Time evolution – left acting case*):

$$\langle \Omega^{(t)} | \overleftarrow{P} = \langle \Omega^{(t+1)} | \tag{4.2.5b}$$

*Proof***:**

$$\langle \Omega^{(t)} | \overleftarrow{P} | j \rangle = \sum_{i',j',i=1}^{r} \omega^{(t)}{}_{i} \langle i | i' \rangle p_{i'j'} \langle j' | j \rangle$$

$$= \sum_{i',j',i=1}^{r} \delta_{i'i} \omega^{(t)}{}_{i} p_{i'j'} \delta_{j'j'} = \sum_{i}^{r} \omega^{(t)}{}_{i} \, p_{ij} = \langle \Omega^{(t+1)} | j \rangle \tag{4.2.6}$$

The last step comes from Eq. (4.2.2c). In general, we have:

$$\langle \Omega^{(n)} | = \langle \Omega^{(0)} | \overleftarrow{P}^{n} \tag{4.2.7}$$

Using PBN, we can express Example 4.1.1 in our row vector and operator:

$$\begin{aligned} &\langle \Omega^{(3)} | = \langle \Omega^{(0)} | \overleftarrow{P}^{3} = \{ \frac{1}{3} \langle 1 | + \frac{1}{3} \langle 2 | + \frac{1}{3} \langle 3 | \} \overleftarrow{P}^{3} \\ &= .401 \langle 1 | + .198 \langle 2 | + .401 \langle 3 | \equiv [.401,\ .198,\ .401] \end{aligned} \tag{4.2.8}$$

We can also check that the left-acting operator $\boldsymbol{P}$ acting on a PCV on its right does not transit the state as expected and does not always produce a valid PCV (not normalized to 1). For example, when the left-acting gives the desired probability vector, the right-acting may make no sense:

$$[1,0,0] \begin{pmatrix} 1/2 & 1/4 & 1/4 \\ 1/2 & 0 & 1/2 \\ 1/4 & 1/4 & 1/2 \end{pmatrix} = [1/2, 1/4, 1/4], \qquad \begin{pmatrix} 1/2 & 1/4 & 1/4 \\ 1/2 & 0 & 1/2 \\ 1/4 & 1/4 & 1/2 \end{pmatrix} \begin{bmatrix} 1 \\ 0 \\ 0 \end{bmatrix} = \begin{bmatrix} 1/2 \\ 1/2 \\ 1/4 \end{bmatrix}$$

The fact that the transition matrix of a Markov chain is left-acting on a PRV is just the result of the conventional definition (4.2.2a). If we define a transition matrix as the transpose of (4.2.2a):

$$p^{T}{}_{ji} \equiv p_{ij} \tag{4.2.9}$$

Then we can build a right-acting transition operator acting on a PCV, already given by Eq. (4.2.3a).

**Example 4.2.2** (The Land of Oz**,** right-acting case)**:** We rewrite the matrix in Example 4.1.1 and Eq. (4.2.8), using their transpose. The matrix and the VCR become:

$$P^T = \begin{pmatrix} 1/2 & 1/2 & 1/4 \\ 1/4 & 0 & 1/4 \\ 1/4 & 1/2 & 1/2 \end{pmatrix} \tag{4.2.10}$$

$$\begin{aligned} |\,\Omega^{(3)}\rangle &= (\hat{P}^T)^3\,|\,\Omega^{(0)}\rangle = (\hat{P}^T)^3\{\frac{1}{3}|\,1\rangle + \frac{1}{3}|\,2\rangle + \frac{1}{3}|\,3\rangle\} \\ &= .401|\,1\rangle + .198|\,2\rangle + .401|\,3\rangle = \begin{bmatrix} .401 \\ .198 \\ .401 \end{bmatrix} \end{aligned} \tag{4.2.11}$$

If the transition matrix is symmetric, then the corresponding operator becomes bidirectional. Such a matrix may have important applications in diffusion maps proposed by Lafon for data clustering [8-9]. Their starting point is the transition matrix of a Markov chain with some special conditions. We find that symmetric transition matrices may provide good examples for diffusion maps. But this is beyond the scope of this article. We leave it to our future work.

# 5. Probability Bracket Notation and Stochastic Processes

## 5.1. Stochastic Processes in the PBN

The Markov chain we have discussed is a special case (discrete-time, discrete-outcome, and homogenous) of the Markov Processes.  Many stochastic processes are also Markov chains. In this section, we want to apply the PBN to some basic formulas of several important stochastic processes [10-14]. We will see that the PBN can simplify the formulation and represent the evolution equation in both the Heisenberg and Schrodinger pictures, as used in QM [15].

**Basic Notations for a Stochastic Process (SP):** The base *P*-ket of a SP $X(t)\,(t \in T)$ is a time-dependent observable and we can generalize Proposition 2.2.1 (observable and its eigen-ket) for discrete or continuous random variables (R.V):

$$\begin{aligned} X(t)|\,X(t) = x_i) &\equiv X(t)|\,x_i, t) = x_i\,|\,x_i, t) \\ P(x_i, t\,|\,x_j, t) &= P(x_i\,|\,x_j) = \delta_{ij} \end{aligned} \qquad \text{(Discrete } R.V.) \tag{5.1.1a}$$

$$X(t)|\,X(t)=x)\equiv X(t)|\,x,t)=x|\,x,t) \qquad \text{(Continuous } R.V.\text{)} \qquad (5.1.1b)$$
$$P(x,t|\,x',t)=P(x|\,x')=\delta(x-x')$$

The time-dependent sample space Ω($t$) contains all possible outcomes observed at time = $t$. Assuming its system P-ket is normalized, we have:

$$P(\Omega\,|\,\Omega(t))=1;\quad P(\Omega|\,x,t)=1;\quad P(x,t)=P(x,t\,|\,\Omega)\equiv P(x\,|\,\Omega(t)) \qquad (5.1.2a)$$

The time-dependent probability distribution now can be written as:

$$P(x_i,t\,|\,\Omega)=P(x_i\,|\,\Omega(t))=m(x_i,t) \qquad (Discrete\, R.V)$$
$$P(x,t\,|\,\Omega)=P(x\,|\,\Omega(t))=f(x,t) \qquad (Continuous\, R.V) \qquad (5.1.2b)$$

The time-dependent expectation value of observable $X(t)$ now is:

$$P(\Omega\,|\,X(t)\,|\,\Omega)=\sum_i P(\Omega\,|\,X(t)\,|\,x_i,t)P(x_i,t\,|\,\Omega)=\sum_i m(x_i,t)\,x_i \quad (Discrete\, R.V)$$
$$P(\Omega\,|\,X(t)\,|\,\Omega)=\int dx P(\Omega\,|\,X(t)\,|\,x,t)P(x,t\,|\,\Omega)=\int dx f(x,t)x \quad (Continuous\, R.V) \qquad (5.1.3)$$

Here we have used the time-dependent identity operator:

$$\hat{I}(t)=\sum_i |\,x_i,t)P(x_i,t\,| \qquad (Disacrete\, R.V)$$
$$\hat{I}(t)=\int dx\,|\,x,t)P(x,t\,| \qquad (Continuous\, R.V) \qquad (5.1.4a)$$

We have a time parameter here because the transition probability or time increment is always defined in the time-incremental direction. When we insert the identity operator, we need to choose the appropriate time (see Eq. (5.1.6)). For example, at time $t$, the measurement picks up the value from $|\,\Omega_t)$ :

$$X_t=X\cdot\hat{I}(t),\quad X(t)\,|\,\Omega)=X\cdot\hat{I}(t)\,|\,\Omega)=X\,|\,\Omega_t) \qquad (5.1.4b)$$
$$P(\Omega\,|\,X_t\,|\,\Omega)=P(\Omega\,|\,X\cdot I(t)\,|\,\Omega)=P(\Omega\,|\,X\,|\,\Omega_t) \qquad (5.1.4c)$$

The shift of the time dependence from the observable to the state P-ket can be thought of as a shift from the Heisenberg picture to the Schrodinger picture (see Eq. (5.2.25) or P.541, [16]).

An S.P. $X(t)$ has independent increments, if for $t_1 < t_2 < \ldots < t_m < t_{m+1}$ then $\forall i\in\{1,\ldots m-1\}$:

$$P(X_{t_{m+1}}-X_{t_m}=x_m\,|\,X_{t_{i+1}}-X_{t_i}=x_{i+1})=P(X_{t_{m+1}}-X_{t_m}=x_m\,|\,\Omega), \qquad (5.1.5a)$$

If an $S.P$ has independent-time increment, we can always set $X_0=0$ and have:

$$P(X_t = x + c \mid X_s = c) = P(X_t - X_s = x \mid X_s - X_0 = c) = P(X_t - X_s = x \mid \Omega) \qquad (5.1.5b)$$

An SP $X(t)$ may have the Markov property, which assumes that the future probability distribution can be predicted from the current system state, but not the past system state. This means, for $t_1 < t_2 < t_{...} < t_m < t_{m+1}$,

$$\begin{aligned} & P(X(t_{m+1}) = x_{m+1} \mid X(t_m) = x_m, X(t_{m-1}) = x_{m-1}, \ldots, X(t_1) = x_1) \\ & \equiv P(x_{m+1}, t_{m+1} \mid x_m, t_m; x_{m-1}, t_{m-1}; \ldots; x_1, t_1) = P(x_{m+1}, t_{m+1} \mid x_m, t_m) \end{aligned} \qquad (5.1.5c)$$

An SP is homogeneous if it has the following property for $t > s > \tau \geq 0$:

$$P([X(t) - X(s) = x] \mid \Omega) = P([X(t-\tau) - X(s-\tau) = x] \mid \Omega) \qquad (5.1.5d)$$

If the SP is homogeneous and $X(0) = 0$, then we have the following property:

$$\begin{aligned} & P([X(t+s) - X(s) = x] \mid \Omega) = P([X(t) - X(0) = x] \mid \Omega) \\ & = P([X(t) = x] \mid \Omega) \equiv P(x \mid \Omega(t)) \end{aligned} \qquad (5.1.5e)$$

The Chapman-Kolmogorov Theorem ([10], p174, p213; [11]-[13]): The equations can be derived by using the Conditional Total Probability Law (CTL) and the Markov property. But we can “derive” them simply by using our identity operators and Eq. (5.1.4):

$$p^{m+n}{}_{ij} \equiv P(j, m+n \mid i, 0) = P(j, m+n \mid I(m) \mid i, 0) = \sum_k P(j, m+n \mid k, m) P(k, m \mid i, 0)$$

$$= \sum_k p^m{}_{ik} \;\; p^n{}_{kj} \qquad (Disacrete time, discrete R.V) \qquad (5.1.6a)$$

$$p_{ij}(t+s) \equiv P(j, t+s \mid i, 0) = P(j, t+s \mid I(s) \mid i, 0) = \sum_k P(j, t+s \mid k, s) P(k, s \mid 0, i)$$

$$= \sum_k p_{ik}(s) \;\; p_{kj}(t) \qquad (Continuous\ time, discrete\ R.V) \qquad (5.1.6b)$$

$$P(x,t \mid y,s) = P(x,t \mid \hat{I}(\tau) \mid y,s) = \int P(x,t \mid z,\tau) dz \, P(z,\tau \mid y,s) \quad where\ \ t > \tau > s$$

$$(Continuous time, continuous R.V) \qquad (5.1.6c)$$

In general, if an *S.P.* has the Markov property, then we can insert an Identity operator (5.1.4a) inside the transition matrix (a P-bracket), with a time less than the time on the left and greater than the time on the right.

Let us list some important examples of stochastic processes.

**Poisson Process** ([10], p.161; [11-13])**:** It is a counting process, $N(t)$, having the following properties:

(1). $\{N(t), t \geq 0\}$ is a non-negative process with independent increments and $N(0) = 0$;
(2). It is homogeneous and its probability distribution is given by:

$$m(k,t) \equiv P([N(t+s) - N(s) = k] \,|\, \Omega) \underset{\substack{independent \\ increament}}{=} P([N(t) - N(0) = k] \,|\, \Omega)$$
$$\underset{N(0)=0}{=} P([N(t) = k] \,|\, \Omega) \equiv P(k \,|\, \Omega(t)) \underset{\substack{Poisson \\ Distribution}}{=} \frac{(\lambda t)^k}{k!} e^{-\lambda t} \tag{5.1.7a}$$

Using the identity operator, one can easily find that:

$$\mu(t) \equiv \bar{N}(t) \equiv P(\Omega \,|\, N(t) \,|\, \Omega) = \sum_k P(\Omega \,|\, N(t) \,|\, k,t) P(k,t \,|\, \Omega)$$
$$= \sum_i k\, m(k,t) = \lambda t; \quad \sigma^2(t) \equiv P(\Omega \,|\, [N(t) - \bar{N}(t)]^2 \,|\, \Omega) = \lambda t \tag{5.1.7b}$$

It can be shown (see [10], p.215; [11], §3, p.1) that the Poisson Process has the Markov property, and its transition probability is:

$$p_{ij}(t) = P([N(t+s) = j] \,|\, N(t) = i) = P([N(t+s) - N(t) = j - i] \,|\, \Omega) = \frac{(\lambda t)^{j-i}}{(j-i)!} e^{-\lambda t}, \text{ if } \; j \geq i$$
$$p_{ij}(t) = 0, \text{ if } \; j < i \tag{5.1.8}$$

**Wiener Process** (see [10], p.159; [11], §8, p.1; [12] §1.1)**:** It is also a homogeneous process $\{W(t), t \geq 0\}$ with independent increments and the condition $W(0) = 0$. Its probability density is a normal distribution $N(0, t\sigma^2)$:

$$f(x,t) \equiv P([W(t+s) - W(s) = x] \,|\, \Omega) \underset{homogeneous}{=} P([W(t) - W(0) = x] \,|\, \Omega)$$
$$\underset{X(0)=0}{=} P([W(t) = x] \,|\, \Omega) \equiv P(x,t \,|\, \Omega) \equiv P(x \,|\, \Omega(t)) \underset{\substack{Normal \\ Distribution}}{=} \frac{1}{\sqrt{2\pi t}\,\sigma} \exp[-\frac{x^2}{2t\sigma^2}] \tag{5.1.9}$$

Using the identity operator, one can easily find that:

$$\tilde{\mu}(t) \equiv P(\Omega \,|\, W(t) \,|\, \Omega) = 0, \quad \tilde{\sigma}^2(t) \equiv P(\Omega \,|\, W(t)^2 \,|\, \Omega) = t\sigma^2 \tag{5.1.10}$$

**Brownian Motion** ([12] §1.3)**:** It is associated with a *standard Wiener process* $W_s(t)$ (with σ =1) as follows:

$$X(t) = X(0) + \mu t + \sigma W_s(t) \tag{5.1.11}$$

Using Eq. (5.10) (with σ = 1), one can easily find that:

$$Drift: \quad \tilde{\mu}(t) \equiv P(\Omega \,|\, [X(t) - X(0)] \,|\, \Omega) = \mu t,$$
$$Variance: \quad \tilde{\sigma}^2(t) \equiv P(\Omega \,|\, [X(t) - \tilde{\mu}(t)]^2 \,|\, \Omega) = t\sigma^2 \tag{5.1.12}$$

If we define:

$$Y(t) = X(t) - X(0) = \mu t + \sigma W_s(t) \tag{5.1.13}$$

Then the probability density $f(y, t)$ of $Y(t)$ is given by:

$$f(y,t)dy \equiv P(y \,|\, \Omega(t))dy \equiv P(y,t \,|\, \Omega)\,dy \equiv P([Y(t) = \mu t + \sigma W(t)] \,|\, \Omega)dy$$
$$= P([W(t) = (y - \mu t)/\sigma] \,|\, \Omega)\,dx = P([x = (y - \mu t)/\sigma], t \,|\, \Omega)dx$$

$$\equiv P([(y - \mu t)/\sigma], t \,|\, \Omega)dx \underset{(5.1.9)}{=} dx \frac{1}{\sqrt{2\pi t}} \exp[-\frac{(y - \mu t)^2}{2t\sigma^2}]$$

$$\because \frac{dx}{dy} = \frac{1}{\sigma}, \therefore f(y,t) \equiv P(y \,|\, \Omega(t)) = \frac{1}{\sqrt{2\pi t}\,\sigma} \exp[-\frac{(y - \mu t)^2}{2t\sigma^2}] \Rightarrow N(t\mu, t\sigma^2) \tag{5.1.14}$$

Therefore, the Brownian motion is just a Wiener process corresponding to a normal distribution $N(t\mu, t\sigma^2)$. Brownian motions also have the Markov property ([13], §1.6).

## 5.2. Time Evolution: The Schrodinger and the Heisenberg Pictures

The time evolution of stochastic processes is an important subject in mathematics, physics, and IR ([8-10]). In this section, we apply our *PBN* to present the Time evolution or the master equation for TC-HMC with discrete state space. We will see that the PBN can make a master equation look like the Schrodinger Equation in QM.

**The Kolmogorov Forward and Backward Equations**: We assume that the Markov chains are stochastically continuous: for infinitesimal $h$, the transition probability has the Tailor expansions ([10], p. 217; [11], §5, p.5; and [14], §6.8):

$$p_{ij}(h) = p_{ij}(0) + p_{ij}'(0)h + o(h^2) = \delta_{ij} + q_{ij}h + o(h^2) \tag{5.2.1}$$

Then, using Eq. (4.1.6b), we have:

$$p_{ij}(t+h) = \sum_k p_{ik}(t)\; p_{kj}(h) = \sum_k p_{ik}(t)\; (\delta_{kj} + q_{kj}h + o(h^2))$$
$$= p_{ij}(t) + \sum_k p_{ik}(t)\; (q_{kj}h + o(h^2)) \tag{5.2.2}$$

Therefore, we get the following Forward equations:

$$p'_{ij}(t) = \lim_{h \to 0} [(p_{ij}(t+h) - p_{ij}(t))/h] = \sum_k p_{ik}(t)\; q_{kj} \tag{5.2.3}$$

Similarly, we can derive the Backward equations:

$$p_{ij}(h+t) = \sum_k p_{ik}(h)\ p_{kj}(t) \Rightarrow p'_{ij}(t) = \sum_k q_{ik}\ p_{kj}(t) \tag{5.2.4}$$

Their matrix forms are:

$$\text{Forward: } P'(t) = P(t)Q; \quad \text{Backward: } P'(t) = QP(t); \tag{5.2.5}$$

They both have the following formal solution with the initial condition $P(0) = I$ :

$$P(t) = P(0)\exp[Qt] = \exp[Qt] = \sum_{k=0}^{\infty} \frac{(Qt)^k}{k!} \tag{5.2.6}$$

As discussed in §4.2, we introduce the transition operator:

$$p_{ij}(t) = P(i\,|\,\hat{P}(t)\,|\,j), \quad q_{ij} = P(i\,|\,\hat{Q}\,|\,j)$$

Then we have the following differential equation ([10], p 220):

$$\frac{d}{dt}\hat{P}(t) = \hat{P}(t)\hat{Q} = \hat{Q}\hat{P}(t); \tag{5.2.7}$$

**The Schrodinger Picture**: We can find the absolute probability ([10], p221) as follows:

$$p_i(t) \equiv P(X(t) = i\,|\,\Omega) \equiv P(i,t\,|\,\Omega) \equiv P(i\,|\,\Omega_t) = \sum_k p_k(0) p_{ki}(t) \tag{5.2.8}$$

It satisfies the following differential equations:

$$\frac{\partial}{\partial t} p_i(t) = \frac{\partial}{\partial t} P(i\,|\,\Omega_t) = \sum_k p_k(t)\, q_{ki} = \sum_k (Q^T)_{ik} P(k\,|\,\Omega_t) = P(i\,|\,Q^T\,|\,\Omega_t) \tag{5.2.9}$$

It is valid for every basis P-bra and leads to the TEDE (or the master equation) in PBN:

$$\frac{\partial}{\partial t}|\,\Omega_t) = Q^T\,|\,\Omega_t) \equiv \hat{L}\,|\,\Omega_t), \quad |\,\Omega_t) = \hat{U}(t)\,|\,\Omega_0) = e^{Q^T t}\,|\,\Omega_0) = e^{\int_0^t dt\hat{L}}\,|\,\Omega_0) \tag{5.2.10}$$

We know that the snapshot of sample space at time *t* can be mapped to a Probability Column Vector (PCV), and can be expanded with a V-base, as expressed in Eq. (4.1.4):

$$|\,\Omega_t\rangle = I\,|\,\Omega_t\rangle = \sum_i^t |\,i\rangle\langle i\,|\,\Omega_t\rangle = \sum_i^t p_i(t)\,|\,i\rangle, \quad |\,\Omega_t\rangle = \hat{U}(t)\,|\,\Omega_0\rangle \tag{5.2.11}$$

This is not new. It is identical to the master equation used in the Doi formalism [16-18] for discrete-state homogeneous Markov chains (like the birth-death process):

$$\frac{\partial}{\partial t}|\psi(t)\rangle = \hat{L}|\psi(t)\rangle, \quad |\psi(t)\rangle = \hat{U}(t)|\psi(0)\rangle = e^{\hat{L}t}|\psi(0)\rangle \tag{5.2.12}$$

The Doi definition of a « **state function** » (see [16] and [18]) can be readily identified as our system state $P$-bra:

$$P(\Omega| = \sum_{\vec{n}} P(\vec{n}| \leftrightarrow \langle s| \equiv \sum_{\vec{n}} \langle \vec{n}|, \quad \therefore \langle \hat{F}(t)\rangle = \langle s|\hat{F}(\vec{n})|\psi(t)\rangle = P(\Omega|\hat{F}(\vec{n})|\Omega_t) \tag{5.2.16}$$

Here, the basis is formed by the eigenvectors of occupation operators in a Fock space:

$$\hat{n}_i|\vec{n}\rangle = n_i|\vec{n}\rangle, \quad \sum_{\vec{n}}|\vec{n}\rangle\langle\vec{n}| = 1, \quad \langle\vec{n}|\vec{n}'\rangle = \delta_{\vec{n},\vec{n}'} = \prod_{i=1}\delta_{n_i,n_i'} \tag{5.2.17}$$

In the Peliti formalism [17], the base (from population $n$) is normalized specifically:

$$\sum_n |n\rangle\frac{1}{n!}\langle n| = \hat{I}, \quad \langle m|n\rangle = n!\delta_{m,n} \tag{5.2.18}$$

Therefore, the system $P$-bra is now expanded as:

$$P(\Omega| = P(\Omega|\hat{I} = \sum_n^{\infty} P(\Omega|n)\frac{1}{n!}P(n| \underset{(2.1.4)}{=} \sum_n^{\infty} P(n|\frac{1}{n!} \tag{5.2.19}$$

Mapping to vector space, it is nothing else, but the « **standard bra** » introduced in [17]:

$$\begin{aligned} &P(\Omega| = \sum_n \frac{1}{n!}P(n| \quad \leftrightarrow \quad \langle| \equiv \sum_n \frac{1}{n!}\langle n|, \\ &\therefore E[\hat{F}] \equiv \langle\hat{F}\rangle = \langle|\hat{F}|\Psi(t)\rangle = P(\Omega|\hat{F}|\Omega_t) \end{aligned} \tag{5.2.20}$$

We call Eq. (5.2.11-12) the master equation in the **Schrodinger picture** because they are similar to the stationary Schrodinger equation of QM in the Dirac (VBN) notation.

$$i\hbar\frac{\partial}{\partial t}|\Psi(t)\rangle = \hat{H}|\Psi(t)\rangle, \quad |\Psi(t)\rangle = \hat{U}(t)|\Psi(0)\rangle = e^{-iHt/\hbar}|\Psi(0)\rangle \tag{5.2.21}$$

**The Heisenberg picture*:*** Note that Eq. (5.2.12) is independent of representations. It is also true for homogeneous MC of continuous states (see App. B). Now we introduce the Heisenberg picture as used in QM ([15], §11.12):

$$|\Omega_t) = \hat{U}(t)|\Omega_0), \quad \Rightarrow \quad \hat{X}(t) = \hat{U}^{-1}(t)\hat{X}\hat{U}(t) \tag{5.2.22}$$

$$\begin{aligned} &\therefore P(\Omega_t|\hat{X}|\Omega_t) = P(\Omega_0|\hat{U}^{-1}(t)\hat{X}\hat{U}(t)|\Omega_0) = P(\Omega_0|\hat{X}(t)|\Omega_0) \\ &= P(\Omega|\hat{X}(t)|\Omega) \end{aligned} \tag{5.2.23}$$

In the last step, we identify $\Omega_0 = \Omega$ in the Heisenberg picture. Based on the definition of $\hat{X}(t)$, we introduce the following relations:

$$\begin{aligned} &| x,t) = \hat{U}^{-1}(t) | x), \quad P(x | \hat{U}(t) = (x,t |, \quad \therefore P(x,t | x',t) = (x | x') \\ &P(x',t | \hat{X}(t) | x,t) = P(x' | \hat{U}(t)\hat{U}^{-1}(t)\, \hat{X}\, \hat{U}(t)\hat{U}^{-1}(t) | x) = P(x' | \hat{X} | x) = x\, P(x' | x) \end{aligned} \tag{5.2.24}$$

Using the fact that $\Omega_0 = \Omega$ in the Heisenberg picture, we get the shift in Eq. (5.1.2):

$$P(x | \Omega_t) = P(x | \hat{U}(t) | \Omega_0) = P(x,t | \Omega_0) = P(x,t | \Omega) = P(x,t) \tag{5.2.25}$$

If the system P-ket $| \Omega_t)$ is normalized, the expectation of observer $X$ reads:

$$\langle \hat{X} \rangle_t = P(\Omega | \hat{X} | \Omega_t) = \mathrm{Tr}[\hat{X} | \Omega_t) P(\Omega |] \tag{5.2.26}$$

Just like in QM, if the system P-ket $| \Omega_t)$ is not normalized, the expectation becomes:

$$\langle \hat{X} \rangle_t = P(\Omega | \hat{X} | \Omega_t) / P(\Omega | \Omega_t) = \mathrm{Tr}[\hat{X} | \Omega_t) P(\Omega |] / \mathrm{Tr}[ | \Omega_t) P(\Omega |] \tag{5.2.27}$$

## 6. Microscopic Probabilistic Processes and Quantum Statistics

Using the PBN with path integrals, we discovered several correlations between the Hilbert space and the probabilistic space by the special Wick rotation (*it*→ *t*) [19], [20]. These correlations enable us to address some special processes of different systems, from one single particle to many identical particles.

### 6.1. The Microscopic Probabilistic Processes of a Single Particle

In Ref. [19], by studying the path integrals, we demonstrated that the stationary Schrodinger equation (5.2.21) in the Hilbert space naturally changes to the master equation (5.2.10) in the probability space under the special Wick rotation (SWR). The master equation describes the time evolution of an induced microscopic diffusion [19], which we call a **macroscopic probabilistic process** (MPP) here.

The **Special Wick Rotation** (SWR) is defined by (see Eq. (3.5-6), [19]):

$$\text{SWR: } it \to t, \quad | \psi(t) \rangle \to | \Omega_t), \quad \langle x_b, t_b | x_a, t_a \rangle \to P(x_b, t_b | x_a, t_a) \tag{6.1.1}$$

Under the SWR, a Schrodinger equation becomes the master equation of an MPP:

$$\frac{\partial}{i\,\partial t} | \Psi(t) \rangle = \frac{-1}{\hbar} \hat{H} | \psi(t) \rangle \overset{it \to t}{\to} \frac{\partial}{\partial t} | \Omega_t) = \frac{-1}{\hbar} \hat{H} | \Omega_t) \tag{6.1.2}$$

In our recently published paper [20], we discussed the basic ideas of the PBN (mostly based on this article), the implication of the SWR (as in [19]), the induced microscopic diffusions (or MPP as we called it here) of a single particle, and its possible application to certain systems with non-Hermitian Hamiltonians (§5 of [20], [21]).

For a system of one particle, from Eq. (6.1.2), the time evolution of the eigenstates of the Hamiltonian has the following correlation under the SWR (see Eq. (115), [20]):

$$|\psi_k(t)\rangle = e^{-i\varepsilon_k t/\hbar}|\psi_k\rangle \overset{it\to t}{\rightarrow} |\psi_k(t)) = e^{-\varepsilon_k t/\hbar}|\psi_k) \qquad (6.1.3)$$

Suppose the MPP has a normalized initial system P-ket:

$$|\Omega_0) = \sum\nolimits_{k\in B}\eta_k\,|\psi_k), \quad P(\Omega\,|\,\Omega_0) = \sum\nolimits_{k\in B}\eta_k = 1 \qquad (6.1.4)$$

It leads to the following unnormalized time-dependent system P-ket,

$$|\Omega_t) = \sum\nolimits_{k\in B}\eta_k e^{-\varepsilon_k t/\hbar}|\psi_k), \quad P(\Omega\,|\,\Omega_t) = \sum\nolimits_{k\in B}\eta_k e^{-\varepsilon_k t/\hbar} \neq 1 \quad \text{for } t > 0 \qquad (6.1.5)$$

Applying Eq. (5.2.27), the expectation of energy has the expression:

$$\langle\hat{H}\rangle_t = P(\Omega\,|\,\hat{H}\,|\,\Omega_t)\,/\,P(\Omega\,|\,\Omega_t) \qquad (6.1.6)$$

Due to the exponential time dependence in Eq. (6.1.5), the basis P-ket of the lowest energy level eventually dominates the system P-ket. Let $\varepsilon_s, s\in B$ be the lowest energy, we have

$$|\Omega_t) = \sum\nolimits_{k\in B}\eta_k e^{-\varepsilon_k t/\hbar}|\psi_k) \overset{t\to\infty}{\rightarrow} \eta_s e^{-\varepsilon_s t/\hbar}|\psi_s) \qquad (6.1.7)$$

Therefore, the MPP behaves like a process that will ultimately jump to its lowest energy level, included in its initial condition (6.1.5) (see the discussion after Eq. (117), [20]):

$$\langle\hat{H}\rangle_t = P(\Omega\,|\,\hat{H}\,|\,\Omega_t)\,/\,P(\Omega\,|\,\Omega_t) = \sum\nolimits_{k\in B}\eta_k\,\mathrm{e}^{-\omega_k t}\,\varepsilon_k \,/ \sum\nolimits_{k\in B}\eta_k\,\mathrm{e}^{-\omega_k t} \overset{t\to\infty}{\rightarrow} \varepsilon_s \qquad (6.1.8)$$

It becomes a pure state_with zero von Newman entropy [22] and sounds like a freezing process. Now we want to ask: is the time (originally an imaginary time) in Eq. (6.1.5) related to temperature? The answer seems to be YES.

### 6.2. Many-Particle Systems and the Wick-Matsubara Relation

For a system of many identical and non-interacting particles, we have the following occupation number P-basis in the probability space (see §2.3), corresponding to the V-basis in the Fock space ([1], §22.1):

$$\hat{n}_i\,|\,n_1, n_2, \ldots) = \hat{n}_i \prod\nolimits_k |\,n_k) \equiv n_i\,|\,\vec{n}) \qquad (6.2.1)$$

$$P(\vec{n}'|\vec{n}) = \prod_i P(n_i'|n_i) \equiv \delta_{\vec{n}',\vec{n}}, \quad \sum_{\vec{n}} |\vec{n})P(\vec{n}| = \prod_i \sum_{n_i} |n_i)P(n_i| = \hat{I}_{\vec{n}} \tag{6.2.2}$$

The system has the following total Hamiltonian (Eq. (1.3), [23]):

$$\hat{\boldsymbol{H}} = \hat{H} - \mu\hat{N} = \sum_i \sum_{n_i} \hat{n}_i(\varepsilon_i - \mu) \tag{6.2.3}$$

Here $\hat{H}$, $\varepsilon_i$ and $\mu$ are the conventional dynamic Hamiltonian, the dynamic energy per particle at the $i^{\text{th}}$ level, and the chemical potential per particle. Based on Eq. (6.1.3), the time evolution of the basis P-ket at the $i^{\text{th}}$ level is:

$$\partial_t |n_i,t) = -[n_i(\varepsilon_i - \mu)/\hbar]|n_i,t), \quad |n_i,t) = e^{-n_i(\varepsilon_i-\mu)t/\hbar}|n_i) \tag{6.2.4}$$

We assume that at $t = 0$, the probability of the system is evenly distributed like:

$$|\Omega_{i,0}) \equiv |\Omega_i) = \sum_{n_i} |n_i), \quad |\Omega_{i,t}) = \sum_{n_i} e^{-n_i(\varepsilon_i-\mu)t/\hbar}|n_i) \tag{6.2.5}$$

It is unnormalized and the expected occupation number of particles at the $i^{\text{th}}$ level reads:

$$n(\varepsilon_i) \equiv \langle\hat{n}_i\rangle = \sum_{n_i} n_i e^{-\varepsilon_i n_i t/\hbar} / \sum_{n_i} e^{-\varepsilon_i n_i t/\hbar} \tag{6.2.6}$$

The expected total number of particles can be expressed as:

$$\langle\hat{N}\rangle = \sum_i \langle\hat{n}_i\rangle == \sum_i \left\{\sum_{n_i} n_i e^{-\varepsilon_i n_i t/\hbar} / \sum_{n_i} e^{-\varepsilon_i n_i t/\hbar}\right\} \qquad 6.2.7)$$

It is quite similar to the distribution function in quantum statistics (see [6], 11.2-6):

$$\langle\hat{N}\rangle = \sum_i \left\{\sum_{n_i} n_i e^{-(\varepsilon_i-\mu)n_i/kT} / \sum_{n_i} e^{-n_i(\varepsilon_i-\mu)/kT}\right\} \equiv \sum_i \sum_{n_i} n_i e^{-(\varepsilon_i-\mu)n_i/kT} / z_i \tag{6.2.8}$$

Here $z_i$ is the partition function of the $i^{\text{th}}$ level. Comparing Eq. (6.2.7) and (6.2.8), we find that the imaginary time in the Wick rotation corresponds to the absolute temperature (the Matsubara formalism [23]), which we call:

The Wick-Matsubara Relation: $$it \xrightarrow{\text{Wick}} t \xrightarrow{\text{Matsubara}} \hbar/kT \tag{6.2.9}$$

For fermions, $n_i \in \{0, 1\}$ (the Fermi-Dirac distribution), all lowest energy states ($\varepsilon \leq \varepsilon_F$) are occupied when $T \to 0$ (or $t \to \infty$). For bosons, $n_i \geq 0$ (the Bose-Einstein distribution), when $T \to 0$ (or $t \to \infty$), all particles go to the ground energy level (the Bose condensation). So the systems (either of bosons or fermions) are eventually frozen at zero temperature with the lowest total energy and zero entropy (no chaos).

The grand partition function is expressed as (see [6], 11.2-6, [24], page 37):

$$Z_G \equiv \sum_{\vec{n}} \langle \vec{n} \,|\, e^{-(\hat{H}-\mu\hat{N})/kT} \,|\, \vec{n} \rangle = \prod_{i=1}^{\infty} \sum_{n_i} e^{-(\varepsilon_i - \mu) n_i /kT} = \prod_{i=1}^{\infty} z_i \qquad (6.2.10)$$

We can write it concisely and it is easy to expand in the PBN:

$$Z_G = P(\Omega \,|\, e^{-(\hat{H}-\mu\hat{N})/kT} \,|\, \Omega) = P(\Omega \,|\, e^{-(\hat{H}-\mu\hat{N})/kT} \hat{I}_{\vec{n}} \,|\, \Omega) = \prod_{i=1}^{\infty} \sum_{n_i} e^{-(\varepsilon_i - \mu)\hat{n}_i /kT} \qquad (6.2.11)$$

The freezing behavior of an MPP implies that the system must be in contact with a huge reservoir with an absolute zero temperature. Since the master equation of an MPP describes a probabilistic process, even though Eq. (6.1.5) is about a single particle, the Wick-Matsubara relation may still be applied.

## Summary

In this article, we proposed a new set of symbols of the PBN (Probability Bracket Notation) in the probability sample space. Using the PBN, various definitions or formulas in the probability theory can be represented and manipulated just like their counterparts in QM using the Dirac notation or VBN (Vector Bracket Notation). We derived the master equation of time-continuous MC, independent of the P-basis. We identified our system P-bra with the state function in the Doi formalism or the standard bra in the Peliti Techniques. We explained the implication of the change from the Schrodinger to the Heisenberg picture. We explored the microscopic probabilistic processes (MPP) and reproduced the basic distribution function in quantum statistics, connecting time with temperature ($t \rightarrow \hbar/kT$). The freezing behavior of an MPP implies that the system must be in contact with a huge reservoir with absolute zero temperature.

Of course, more investigations are needed to verify the PBN's consistency (or correctness), usefulness, and limitations. We did some additional work. For example, in Ref [25], we discussed the systems of multiple random variables (dependent or independent) and the introduction to static Bayesian networks.

### Appendix A: The Comparison of the PBN and the VBN

**Table 1: Probability Bracket vs. Vector Bracket (discrete case)**

| | ***PBN*** | ***VBN*** |
|---|---|---|
| Space | Sample space $\Omega$, associated with a random variable $X$ | Hilbert space $\boldsymbol{H}$, associated with a Hermitian operator $H$ |
| Bra | $P(A\|$: an event set in $\Omega$<br>$P(\Omega\|$: state $P$-bra | $\langle \psi_A \|$: a (row) vector in $\boldsymbol{H}$<br>$\langle \Psi(t) \|$: V-state bra |

| Ket | $\vert B)$: an evidence set in $\Omega$<br>$\vert \Omega_t)$: state $P$-ket | $\vert \psi_B\rangle$: a (column) vector in $\boldsymbol{H}$<br>$\vert \Psi(t)\rangle = \langle \Psi(t)\vert^\dagger$ V-state ket |
|---|---|---|
| Bracket | $P(A/B)$: $P$-bracket<br>(Conditional probability) | $\langle \psi_A/\psi_B\rangle \equiv \langle \psi_A,\ \psi_B\rangle$: V-bracket<br>(Inner vector product) |
| Bracket transpose | Using Bayes formula | $\langle \psi_B/\psi_A\rangle = \langle \psi_A/\psi_B\rangle^*$ |
| Special relations<br>(*PBN* only) | $P(A/B) = 1$ if $A \supseteq B \supset \varnothing$<br>$P(A/B) = 0$ if $A \cap B = \varnothing$<br>$P(A \mid B) = P(A \mid \Omega)$, if $A$ and $B$ are mutually indepedent | |
| Base origin | Complete mutual-disjoint sets associated with variable $X$:<br>$\omega_i \cap \omega_j = \delta_{ij}\, \omega_i$, $\Sigma \omega_i = \Omega$ | Eigenvectors of a Hermitian Operator $H$:<br>$\hat{H} /\psi_i\rangle = E_i /\psi_i\rangle$ |
| Orthonormality | $P(\omega_i / \omega_j) = \delta_{ij}$ | $\langle \psi_i/\psi_j\rangle = \delta_{ij}$ |
| Identity operator | $\hat{I} = \Sigma \vert \omega_i) P(\omega_i \vert$ | $\hat{I} = \Sigma /\psi_i\rangle\langle \psi_i \vert$ |
| Right expansion | $\vert \Omega_t) = \Sigma /\omega_i) P(\omega_i \vert \Omega_t)$<br>$= \Sigma\, m_i(t) /\omega_i)$ | $\vert \Psi(t)\rangle = \Sigma \vert \psi_i\rangle\langle \psi_i \vert \Psi(t)\rangle$<br>$= \Sigma\, c_i(t) \vert \omega_i)$ |
| Left expansion | $P(\Omega \vert = \Sigma\, (\Omega \vert \omega_i) P(\omega_i \vert$<br>$= \Sigma\, P(\omega_i \vert$ | $\langle \Psi(t) \vert = \vert \Psi(t)\rangle^\dagger$ |
| State normalization | $P(\Omega \vert \Omega_t) = \Sigma\, m_i(t) = 1$ | $\langle \Psi(t) \vert \Psi(t)\rangle = \Sigma \vert c_i(t) \vert^2$ |
| Observable | $X/\omega_i) = x_i/\omega_i)$ | $\hat{H} /\psi_i\rangle = E_i /\psi_i\rangle$ |
| Expectation value | $\langle X\rangle = P(\Omega \vert X \vert \Omega_t) = \Sigma\, m_i(t)\, x_i$ | $\langle H\rangle = \langle \Psi(t) \vert\ \hat{H} \vert \Psi(t)\rangle$<br>$= \Sigma \vert c_i(t) \vert^2 E_i$ |

**Table 2: Probability Bracket vs. Vector Bracket (continuous case)**

| | ***PBN*** | ***VBN*** |
|---|---|---|
| Special relations<br>(*PBN* only) | $P(A \mid B) = 1$ if $A \supset \bar{a} \supset B \neq \varnothing$ & $\int_{\bar{a}} dx > 0$,<br>$P(A / B) = 0$ if $A \cap B = \varnothing$,<br>$P(x \mid x') = \delta(x - x')$ if $x$ and $x'$ are basis events $\in \Omega$<br>$P(A \mid B) = P(A \mid \Omega)$, if $A$ and $B$ are mutually indepedent | |

| Observable density/ distribution Function | $X \mid x) = x \mid x)$<br>$f(x, t) = P(x \mid \Omega_t)$ | $\hat{p} \mid \psi_p\rangle = p \mid \psi_p\rangle$<br>$c(p, t) = \langle p \mid \Psi(t)\rangle$ |
|---|---|---|
| Orthonormality | $P(x \mid x') = \delta(x - x')$ | $\langle \psi_p \mid \psi_{p'}\rangle = \delta(p - p')$ |
| Identity operator | $\hat{I} = \int \mid x)\, dx\, P(x \mid$ | $\hat{I} = \int \mid \psi_p\rangle\, dp\, \langle \psi_p \mid$ |
| Right expansion | $\mid \Omega_t) = \int \mid x)\, dx\, P(x \mid \Omega_t)$<br>$= \int \mid x)\, f(x, t)\, dx$ | $\mid \Psi(t)\rangle = \int \mid \psi_p\rangle\, dp\, \langle \psi_p \mid \Psi(t)\rangle$<br>$= \int \mid p\rangle\, c(p, t)\, dp$ |
| Left expansion | $P(\Omega \mid = \int P(\Omega \mid x)\, dx\, (x \mid$<br>$= \int dx\, P(x \mid$ | $\langle \Psi(t) \mid = \mid \Psi(t)\rangle^{\dagger}$ |
| State normalization | $P(\Omega \mid \Omega_t) = \int f(x, t)\, dx = 1$ | $\langle \Psi(t) \mid \Psi(t)\rangle = \int \mid c(p, t)\mid^2 dp = 1$ |
| Expectation value | $\langle X\rangle = P(\Omega \mid X \mid \Omega_t)$<br>$= \int f(x, t)\, x\, dx$ | $\langle \hat{p} \rangle = \langle \Psi(t) \mid \hat{p} \mid \Psi(t)\rangle$<br>$= \int \mid c(p, t)\mid^2 p\, dp$ |

## Appendix B: The Derivation of the Master Equation of Homogeneous Markov Chains (HMC) with Continuous States

Eq. (5.2.9) and (5.2.10) are representation-independent. They can be easily extended to the homogeneous *M.C* of continuous states. Let us assume that the system is in the $i^{\text{th}}$ state and it is located in the range of $(x, x+\Delta x)$, hence $P(i \mid \Omega(t)) \to P(x \mid \Omega(t))\Delta x$ and:

$$\frac{\partial}{\partial t} P(i \mid \Omega_t) \to \frac{\partial}{\partial t} \Delta x_i P(x_i \mid \Omega_t) = \sum\nolimits_j \Delta x_i P(x_i \mid \hat{Q}^T \mid x_j) \Delta x_j P(x_j \mid \Omega_t)$$

$$\underset{\Delta x \to 0}{\to} \frac{\partial}{\partial t} P(x \mid \Omega_t) = \int dx' P(x \mid \hat{Q}^T \mid x') P(x' \mid \Omega_t) = P(x \mid \hat{Q}^T \mid \Omega_t)$$

$$\to \frac{\partial}{\partial t} P(x,t) = \int dx' P(x \mid \hat{Q}^T \mid x') P(x',t) \equiv \int dx' L(x,x') P(x',t) \qquad \text{(B.1)}$$

We can also derive Eq. (B.1) from the following master equation (Eq. (4.6) of [12]):

$$\frac{\partial}{\partial t} P(x,t) = \int dx' [W_t(x \mid x') P(x',t) - W_t(x' \mid x) P(x,t)] \qquad \text{(B.2)}$$

For an M.C of discrete states, the master equation reduces to (see Eq. (4.7) of [12]):

$$\frac{\partial}{\partial t} P_n(t) = \sum\nolimits_{n' \neq n} [W_{nn'}(t) P_{n'}(t) - W_{n'n}(t) P_n(t)] \qquad \text{(B.3)}$$

Because $W_{ij}$ is time-independent (HMC), similar to Eq. (5.2.3), we define:

$$v_i = \lim_{h\to 0}\frac{1-P_{ii}(h)}{h} = \sum_{j\neq i} w_{ji}, \quad w_{ij} = \lim_{h\to 0}\frac{1-P_{ij}(h)}{h} \quad (i \neq j) \tag{B.4}$$

$$w_{ii} = 0, \quad L_{ij} = w_{ij} - v_i\delta_{ij} = \begin{cases} w_{ij} & (i \neq j) \\ -v_i & (i = j) \end{cases} \tag{B.5}$$

Then Eq. (B.3) can be rewritten in the following form:

$$\frac{\partial}{\partial t} P_n(t) = \sum_{n'}[W_{nn'}P_{n'}(t) - v_n\delta_{nn'}P_n(t)] \equiv \sum_{n'} L_{nn'}P_{n'}(t) \tag{B.6}$$

Extending it to a continuous states HMC, we introduce:

$$v(x) = \int_{x'} dx'\, w(x'\,|\, x) \tag{B.7}$$

Then the master equation for the HMC has the same form as in Eq. (B-1):

$$\frac{\partial}{\partial t} P(x,t) = \int dx'[w(x\,|\,x') - \delta(x-x')v(x)]P(x',t) = \int dx'\, L(x,x')P(x',t) \tag{B.8}$$

Using PBN, combining Eq. (5.2.10) and Eq. (B.8) we have obtained the master equation for a continuous-time HMC on either a discrete or continuous basis:

$$\frac{\partial}{\partial t} |\, \Omega_t) = \hat{L}\,|\, \Omega_t) \tag{B.9}$$

$$|\, \Omega_t) = \hat{U}(t,0)\,|\, \Omega_0) = \exp[\hat{L}t]\,|\, \Omega_0) \tag{B.10}$$